\documentclass[preprintnumbers,showpacs,amsmath,amssymb,floatfix,9pt,prd,onecolumn,
superscriptaddress,nofootinbib]{revtex4}
\usepackage{latexsym}
\usepackage{epsfig}
\usepackage{amssymb}
\begin{document}

\title{Compact star in $f(T)$ gravity with Tolman-Kuchowicz metric potential }

\author{Piyali Bhar}
\email{piyalibhar90@gmail.com}
\affiliation {Department of Mathematics, Government General Degree College Singur, Hooghly ,
 West Bengal 712409, India}

\begin{abstract}
Employing $f(T)$ gravity, where $T$ is the torson, we have developed a new model of an anisotropic compact star in this work. Tolman-Kuchowicz (TK) metric potential has been used to solve the set of field equations. Furthermore, the matching conditions for interior and exterior geometry have been discussed. We have considered observation data of the compact star LMC X-4  and analyzed thermodynamical properties (density, pressure, equation of state parameter, square speed of sound, and equilibrium condition) analytically and graphically to test the validity of the solution. The compact star is found to meet the energy conditions. Through the causality condition and Herrera's cracking concept, the stability analysis of the present model has been presented and it confirms the physical acceptability of the solution. It has been shown that the obtained interior solutions for compact stars are consistent with all necessary physical criterions and therefore relevant as well as physically acceptable.

\end{abstract}

\maketitle







\section{Introduction}
The foundation of general relativity is based on the idea that the gravitational field is produced by intrinsic curvature. Alternative, much less popular, but equally relevant explanation of gravity is based on torsion without the existence of curvature. This theory is referred to as the teleparallel equivalent of general relativity (TEGR). TEGR is obtained from an action produced linearly from the torsion scalar with the tetrad being the degree of freedom rather than an action constructed linearly in the Ricci scalar with the metric as the degree of freedom. The Teleparallel Equivalent of General Relativity (TEGR), a different native geometrical explanation of gravity developed by Albert Einstein, employs absolute parallelism and the tangent space as a frame of reference \cite{israelit1985einstein,hayashi1979new}. At each point in tangent space, an orthogonal tetrad is built in TEGR. This tetrad field is used as a dynamical variable of TEGR \cite{aldrovandi2008physics}. The impact of gravity in TEGR is described by the torsion in space-time. The torsion tensor can be used to determine the torsion created in space-time. As a result, the torsion scalar which characterizes the impact of gravity, and the tetrad field which serves as the dynamical variables of the theory are used to develop the field equations of TEGR. The action of TEGR can be modified to provide $f(T)$ gravity by employing a more extensible torsion scalar function \cite{Bahamonde:2021gfp}.

It has been established that $f(T)$ gravity is a preferable hypothesis to explain both cosmic expansion and galactic dynamics. Li et al. \cite{li2018vol} provides a methodical way to explain torsion gravity by using the effective field theory. Additionally, the gravitational wave propagation bound from GW170817 and GRB170817A have previously demonstrated the effectiveness of this method of effective field theory description and usage of specific operators involving torsional terms by the observational requirement \cite{Cai:2018rzd}. Authors have recently looked into the combined interpretation of the tensions $H_0$ and  $\sigma_8$ \cite{yan2020interpreting}. By utilizing the more general torsion scalar function, $f(T)$ gravity is constructed. Other alternative theories of modified gravity have also been established by physicists including $f(R,\,\mathbb{T})$ gravity, where $\mathbb{T}$ is the trace of the energy momentum tensor \cite{harko2011sd}. $f(G)$, and $f(R,\,G)$ (where $G$ is Gauss-Bonnet invariant) gravities were also studied in refs. \cite{bamba2010finite,Nojiri:2005jg}. We shall concentrate on $f(T)$ gravity in the present article.

When considering possible candidates for modified gravity theories, the $f(T)$ gravity is not only thoroughly studied as its $f(R)$ counterpart, but also attention has increased significantly in recent decades. In the field of cosmology, many research works have been performed in $f(T)$ gravity which can be found in refs. \cite{Rodrigues:2012qua,Salako:2013gka,Paliathanasis:2016vsw,de2016model}.

In the references \cite{Boehmer:2012uyw, Ferraro:2011ks, Astashenok:2020cqq, Pretel:2021kgl}, compact stellar objects like as neutron stars, strange stars and black holes are described using teleparallel gravity. To model compact stars researchers generally use the isotropy pressure. However, new investigations of compact stellar objects like 4U 1820-30, 4U 1538-52, and PSRJ 1614-2230 have shown that the interior of the stellar structures contain anisotropic matter distributions. Such stars have an anisotropic distribution of matter due to the presence of pion condensations, phase transitions, type III-A super fluids etc.
Anisotropic compact stars with diagonal and off-diagonal tetrads in the context of $f(T)$ gravity can be found in \cite{Bahamonde:2021gfp}.
Farrugia et al. \cite{Farrugia:2018gyz} investigated the gravitational waves and their properties under a variety of modified teleparallel theories, such as the $f(T)$, $f(T,B)$, and $f(T,T_G)$ gravities. In this paper the authors employed both the metric and the tetrad languages for the sake of clarity when executing the perturbation analysis around a Minkowski background in the presence of the cosmological constant. Using modified $f(T)$ gravity, Solanki and Said \cite{Solanki:2022yna} created a new class of analytical solutions that describes the anisotropic stellar structures of observed neutron stars. In order to analyze spherically symmetric objects in $f(T)$ gravity, the off-diagonal tetrad is used in the aforementioned work. They introduced physically acceptable metric potentials that can adequately characterize a wide range of astrophysical systems in order to establish exact solutions in the quadratic model of $f(T)$ gravity. In $f(T)$ theory of gravity, Ditta et al. \cite{Ditta:2021wfl} investigated anisotropic star structures with quintessence by using an embedding approach. For a  collection of non-diagonal tetrads in $f(T)$ gravity, Daouda et al. \cite{Daouda:2012nj} investigated the properties of anisotropic fluid. Analytical models of anisotropic strange stars in $f(T)$ gravity with off-diagonal tetrad were produced by Zubair and Abbas \cite{Zubair:2015cpa}.
Using the perturbative method, Nasheda and Saridakis \cite{Nashed:2021pah} investigated at the stability and the thermodynamics properties of spherically symmetric solutions in $f(T)$ gravity. Motivated by all the previous work done, in the present paper we propose a new class of compact stars in $f(T)$ gravity by using pressure anisotropy.\\

Our paper is structured as follows: in Section \ref{2}, we introduce the basic ideas underlying teleparallel gravity and briefly discuss the associated terms which are necessary for constructing the theory of teleparallel gravity. The action of $f(T)$ gravity is then varied in order to get the field equations in $f(T)$ gravity. In Section \ref{3}, we use TK metric potentials to solve the field equations of $f(T)$ gravity. We match the interior metric potential of the star with its exterior Schwarzschild metric in this section. We describe the physical analysis of the model in Section \ref{4}. For various values of the parameter $\alpha$, numerous physical properties of the star including matter density and pressures, their radial derivative, the sound speed etc. are computed and graphically displayed. Numerical values of model parameters like central density, surface density, central pressure etc. for the compact star are presented in tabular form. All the findings from this paper are summarized in Section \ref{last}.

\section{Interior spacetime and Einstein's field Equation in $f(T)$ theory}\label{2}
Generally, one starts with two assumptions when describing relativistic compact star in a theory of gravity: the spherically symmetric static space-time metric given by
\begin{equation}\label{line}
ds^{2}=e^{\nu(r)}dt^{2}-e^{\lambda(r)}dr^{2}-r^{2}(d\theta^2+\sin^2\theta d\phi^2),
\end{equation}
and the matter inside it with a diagonal energy momentum tensor. For our present paper we assume that the matter inside the fluid sphere is anisotropic in nature and its energy momentum tensor is given by,
\begin{eqnarray}
T^{\mu}_{\nu}= diag(\rho,\,-p_r,\,-p_t,\,-p_t),
\end{eqnarray} where $\rho$ , $p_r$ and $p_t$ are the energy density, radial and transverse pressure respectively. In the present work, $\rho$, $p_r$, $p_t$ and the metric functions $e^{\nu}$ and $e^{\lambda}$ are taken to be independent of time. Therefore these two metric functions can only be expressed as functions of $r$ because of spherical symmetry. In our present discussion we use the geometric unit i.e., $G=1=c$. \par
Now, this line element (\ref{line}) can be transformed to a Minkowskian space via the matrix transformation known as the tetrad as follows:
\begin{eqnarray}
ds^2&=&g_{\mu \nu}dx^{\mu}dx^{\nu}=\eta_{ij}\theta^i\theta^j,\\
dx^{\mu}&=&e_i^{\mu}\theta^i,~\theta^i=e^i_{\mu}dx^{\mu}.  \label{tet}
\end{eqnarray}
where
$\eta_{ij}=diag(1,-1,-1,-1)$ and
$e_{i}^{\mu}e^{i}_{\nu}=\delta^{\mu}_{\nu}$. The determinant of the metric $g$ is connected to the determinant of the tetrad
$e=\sqrt{-g}=det(e^{i}_{\mu})$.\\
The components of the torsion and the contorsion  are respectively defined as \cite{HamaniDaouda:2011iy}
\begin{eqnarray}
T^{\sigma}_{\mu\nu}&=&\Gamma^{\sigma}_{\nu\mu}-\Gamma^{\sigma}_{\mu\nu}=e^{\sigma}_{i}\left(\delta_{\mu}e^{i}_{\nu}-\delta_{\nu}e^{i}_{\mu}\right),\label{e1}\\
K^{\mu\nu}_{\sigma}&=&-\frac{1}{2}\left(T^{\mu\nu}_{~~\sigma}-T^{\nu\mu}_{~~\sigma}-T^{\mu\nu}_{\sigma}\right)\label{e2},
\end{eqnarray}
In order to construct the following new tensor given by $S^{\mu\nu}_{\sigma}$, the previous two tensors given by eqs. (\ref{e1}) and (\ref{e2})
are used and the components of the tensor $S_{\sigma}^{\mu \nu}$ is described as,
\begin{equation}
S^{\mu\nu}_{\sigma}=\frac{1}{2}\left(K^{\mu\nu}_{~~\sigma}+\delta^{\mu}_{\sigma}T^{\beta\nu}_{~~\beta}-\delta^{\nu}_{\sigma}T^{\beta\mu}_{~~\beta}\right).
\end{equation}
The torsion scalar is obtained from torsion and contorsion scalar are as follows:
\begin{equation}
T=S^{~~\mu\nu}_{\sigma}T^{~~\sigma}_{\mu\nu}.
\end{equation}
Now the action of $f(T)$ theory is given as \cite{Bengochea:2008gz,Li:2010cg}:
\begin{equation}
S[e^i_{\mu},\phi_A]=\int{d^4x~e~\left[\frac{1}{16\pi}f(T)+\mathcal{L}_{\text{matter}}(\phi_A)\right]},
\end{equation}
Here, $\phi_A$ denotes matter fields and $f(T)$ is an arbitrary analytic function of the torsion scalar $T$.


By varying the action with respect to the tetrads, the field equations of $f(T)$ gravity can be obtained as \cite{Li:2010cg}
\begin{equation}
S^{~~\nu\rho}_{\mu}\partial_{\rho}T f_{TT}+\Big[e^{-1}e^i_{\mu}\partial_{\rho}(e e^{~~\alpha}_iS^{~~\nu \rho}_{\alpha})\Big]f_{T}
+\frac{1}{4}\delta^{\nu}_{\mu}f=4\pi T^{\nu}_{\mu},
\end{equation}
where
\begin{equation}
f_T=\frac{\partial
f}{\partial T}~~~~f_{TT}=\frac{\partial^2f}{\partial T^2}.
\end{equation}
We re-write the line element (\ref{line}) into the invariant form under the Lorentz transformations by defining the tetrad matrix (\ref{tet}) as
\begin{equation}
e^{i}_{\mu}=\begin{pmatrix}
              e^{\frac{\nu}{2}} & 0 & 0 & 0 \\
              0& e^{\frac{\lambda}{2}} & 0 & 0 \\
              0 & 0 & r& 0 \\
              0 & 0 & 0 & r\sin\theta \\
            \end{pmatrix},
\end{equation}
and $e=det(e^i_{\mu})=e^{\frac{\lambda+\nu}{2}} r^2 \sin \theta$
and correspondingly the torsion scalar and its derivative is obtained as,
\begin{eqnarray}
T(r)&=&\frac{2e^{-\lambda}}{r}\left(\nu'+\frac{1}{r}\right),\\
T'(r)&=&\frac{2e^{-\lambda}}{r}\left(\nu''-\frac{1}{r^2}\right)-T\left(\lambda'+\frac{1}{r}\right).
\end{eqnarray}
The field equations in $f(T)$ gravity for anisotropic fluid are given by,
\begin{eqnarray}
4\pi \rho &=&\frac{f}{4}-\left[T-\frac{1}{r^2}-\frac{e^{-\lambda}}{r}(\lambda'+\nu')\right]\frac{f_T}{2}-\frac{e^{-\lambda}}{r}T'f_{TT}, \label{field1}\\
4\pi p_r&=&\left[T-\frac{1}{r^2}\right]\frac{f_T}{2}-\frac{f}{4}, \label{field2}\\
4\pi p_t&=&\left[\frac{T}{2}+e^{-\lambda}\left\{\frac{\nu''}{2}+\left(\frac{\nu'}{4}+\frac{1}{2r}\right)(\nu'-\lambda')\right\}\right]\frac{f_T}{2}-\frac{f}{4}
+\frac{e^{-\lambda}}{2}\left(\frac{\nu'}{2}+\frac{1}{r}\right)T'f_{TT}, \label{max2}
\end{eqnarray}

Here $\rho,~p_r$ and $p_t$ respectively denote the matter density, radial and transverse pressure of the anisotropic fluid sphere and `prime' denotes differentiation with respect to the radial co-ordinate `r'. \\
The fact that the aforementioned field equations (\ref{field1})-(\ref{max2}) lead to the corresponding field equations in GR for $f (T) = T$. In contrast to GR, there is an additional non-diagonal equation in $f(T)$ gravity as follows :
\begin{eqnarray}
e^{-\frac{\lambda}{2}}\frac{\cot\theta}{2r^2}T'f_{TT}&=&0. \label{tor}
\end{eqnarray}

The situations that correspond to eq. (\ref{tor}) is (i) $T'=0$ or (ii) $f_{TT} = 0$. In the second case, one obtains a linear form of the $f(T)$ function as \begin{eqnarray}
f(T)=\alpha T+\beta,
\end{eqnarray}
where $\alpha$ and $\beta$ are constants of integration.\\
It might be argued that one can choose $T' = 0$ instead of $f_{TT} = 0$. If we choose $T'=0$, it gives constant torsion scalar, $T = T_0$. However, as a relativistic star solution, such a solution would be absurd. If one examines the equation (\ref{field2}), it is simple to realize this. As `r' goes to zero for a constant $T$, the right side of this equation blows up. This implies that the pressure of the substance would also blow. This type of solution is absurd and hence we exclude this case \cite{Deliduman:2011ga}.

\section{Solution of the present model and boundary condition}\label{3}

To solve the field equations we choose the metric potentials $e^{\nu(r)}$ and $e^{\lambda(r)}$ are of Tolman-Kuchowicz (TK) type, i.e., we selected as $\nu(r) = Br^2 + 2 \ln D$ and $\lambda(r) = \ln(1 + ar^2 + br^4)$, respectively. Here, the arbitrary constants $a,\, b,\, B$, and $D$ can be evaluated from matching condition. These potentials exhibit good behavior, meet requirements for physical acceptance and are not affected by the central singularity. In the context of modified $f(R,G)$ gravity, Javed et al. \cite{Javed:2021xug} reviewed a few anisotropic stellar spheres. They took the Tolman-Kuchowicz spacetime for this purpose. Anisotropic matter distribution was taken into consideration by the authors when formulating the equations of motion.
In the context of massive Brans-Dicke gravity, Majid and Sharif \cite{Majid:2020hlg} developed an anisotropic model that highlights key characteristics of strange stars. By including the MIT bag model and applying Junction conditions to the boundary of the stellar model, they developed the field equations for the Tolman-Kuchowicz ansatz and estimate the unknown constants in terms of the mass and radius of the star. The current study by Naz and Shamir \cite{Naz:2020ncs} emphasises the impact of electric charge on static spherically symmetric star models in the presence of anisotropic matter distribution under the modified $f(G)$ gravity model. They specifically took into account the stability requirements and singularity-free metric potentials of Tolman-Kuchowicz space-time for this purpose. In order to study an anisotropic spherically symmetric strange star in the background of $f(R,\,T)$ gravity, Biswas et al. \cite{Biswas:2020gzd} used metric potentials of the Tolman-Kuchowicz type that are well-behaved, stable and singularity-free. In order to solve the field equations, Bhar et al. \cite{Bhar:2019lbw} suggested the study of anisotropic compact matter distributions within the context of five-dimensional Einstein-Gauss-Bonnet gravity. They took into account that the inner geometry is described by Tolman-Kuchowicz spacetime.\\

\begin{figure}[htbp]
    \centering
        \includegraphics[scale=.45]{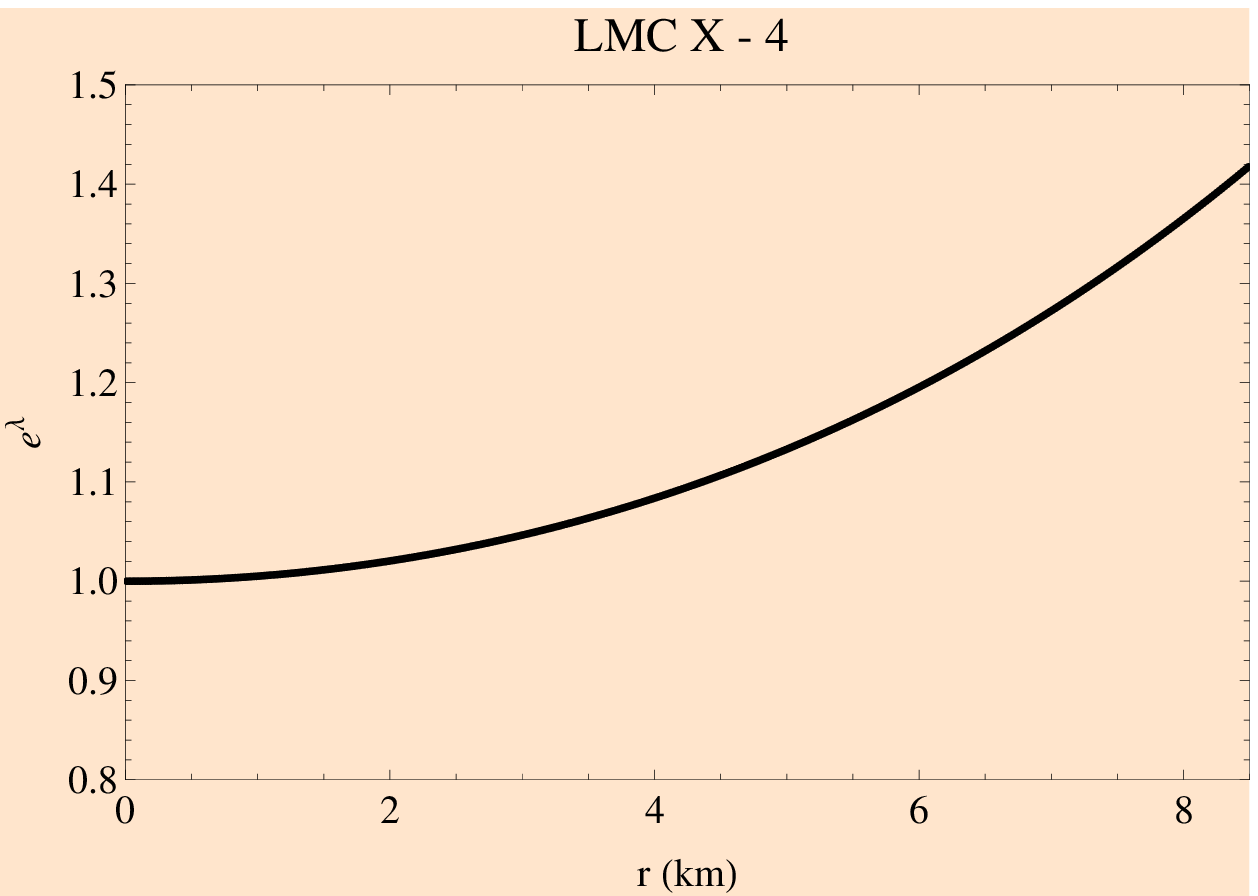}
        \includegraphics[scale=.45]{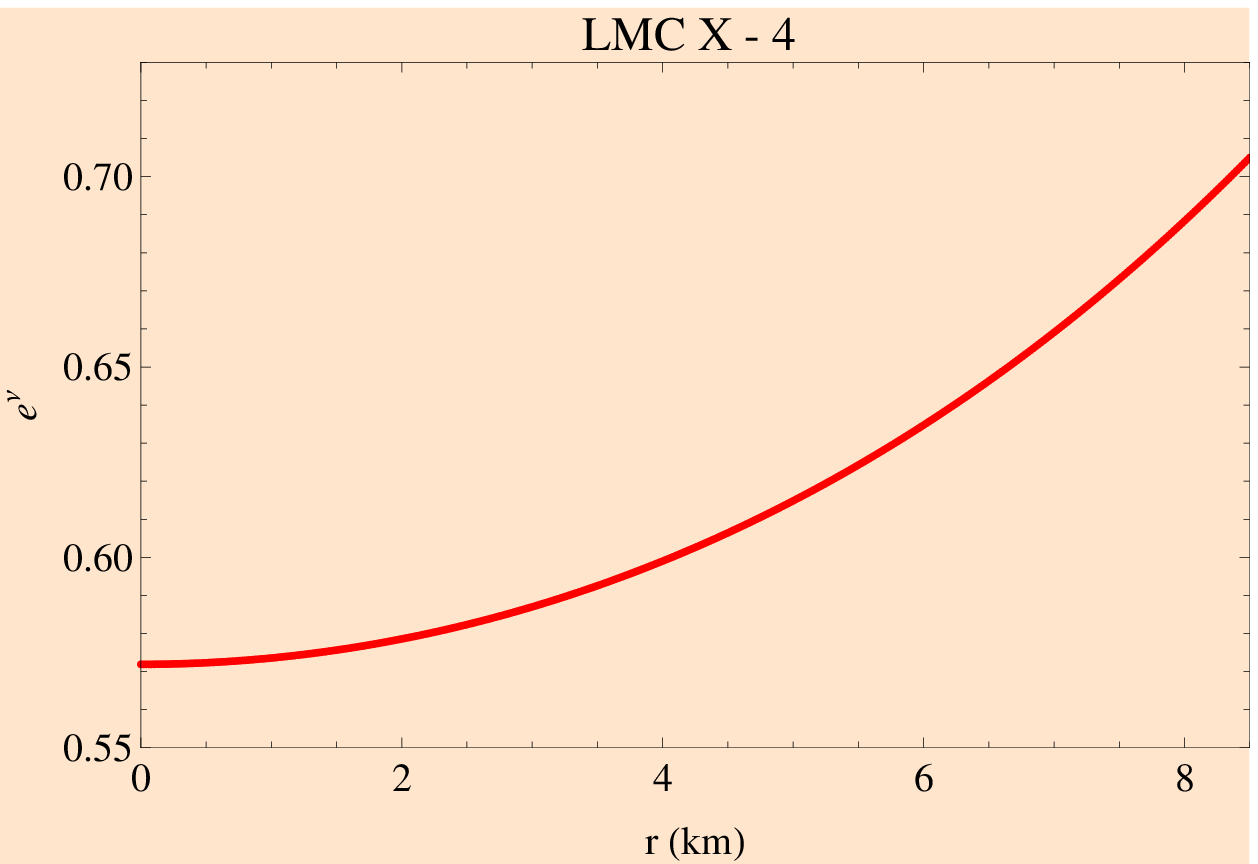}
       \caption{The metric potentials are shown against `r'}
    \label{metric}
\end{figure}

Using these metric potentials, we solve the field equations and obtained the expression of the model parameters as follows:

\begin{eqnarray}
\rho&=&\frac{1}{16\pi} \left[\beta + \frac{
   2 \alpha (3 a + (a^2 + 5 b) r^2 + 2 a b r^4 + b^2 r^6)}{(1 + a r^2 +
      b r^4)^2}\right],\\
      p_r&=&  \frac{\alpha ( 2 B -a- b r^2)}{8\pi (1 + a r^2 + b r^4)}-\frac{\beta}{16\pi},\\
      p_t&=&\frac{1}{16\pi} \left[-\beta + \frac{
   2 \alpha \left(-a + 2 B + (-2 b + B (a + B)) r^2 + a B^2 r^4 +
      b B^2 r^6\right)}{(1 + a r^2 + b r^4)^2}\right].
\end{eqnarray}
The behavior of density and pressure are discussed in the next section.\par
The schwarzschild spacetime given by
\begin{eqnarray}\label{ex}
ds_+^{2}&=&-\left(1-\frac{2M}{r}\right)dt^{2}+\left(1-\frac{2M}{r}\right)^{-1}dr^{2}+r^{2}\left(d\theta^{2}+\sin^{2}\theta d\phi^{2}\right),
\end{eqnarray}
is used as exterior spacetime for our present solution as we have examined uncharged compact stars. The first and second fundamental forms should be imposed across the boundary in order to join the interior spacetime with the exterior Schwarzschild metric and to determine the fundamental constants. The intrinsic metric on $\Sigma$ is continuous according to the first fundamental form which can be written as follows :\\
At the boundary $r=R$, $g_{rr}^+=g_{rr}^-,\, ~~\text{and} ~~~~g_{tt}^+=g_{tt}^-,$\\
where ($-$) and ($+$) sign respectively denotes interior and exterior spacetime. The above two relationships imply,
\begin{eqnarray}
\left(1-\frac{2M}{R}\right)^{-1}&=&1 + aR^2 + bR^4,\label{o1}\\
1-\frac{2M}{R}&=& D^2 e^{BR^2},\label{o2}
\end{eqnarray}
also at $r=R$ : $\frac{\partial}{\partial r}(g_{tt}^+)=\frac{\partial}{\partial r}(g_{tt}^-)$, that gives,
\begin{eqnarray}\label{o4}
\frac{M}{R^3}=BD^2e^{BR^2}.
\end{eqnarray}
The continuity of extrinsic curvature across the boundary yields $p_r(R) = 0$ which gives,
\begin{eqnarray}\label{o3}
\frac{\alpha ( 2 B -a- b R^2)}{8\pi (1 + a R^2 + b R^4)}-\frac{\beta}{16\pi}=0.
\end{eqnarray}
The radius $r = R$ which signifies the size of the compact star is determined by the aforementioned criteria.
Equations (\ref{o1})-(\ref{o3}) are solved to yield the following values for the model parameter:
\begin{eqnarray*}
a &=& \frac{1}{R^2}\left[\left(1 - 2\frac{M}{R}\right)^{-1} - 1 - bR^4\right],\\
B &=& \frac{M}{R^3}\left(1 - 2\frac{M}{R}\right)^{-1},\\
 D &=& e^{-\frac{B R^2}{2}}\sqrt{1 - 2\frac{M}{R}},\\
\beta&=&-\frac{2 \alpha (a - 2 B + b R^2)}{1 + a R^2 + b R^4}.
\end{eqnarray*}
One can note that all the model parameters are successfully obtained in terms of mass and radius of the compact star.
\section{Physical Analysis}\label{4}
We examine a number of physical attributes of anisotropic dense stellar objects in this section including energy density, radial and transversal pressure, anisotropic factor, mass function, compactification factor, surface redshift, adiabatic index etc. for various values of $\alpha$, for a fixed compact star both analytically and graphically.
\subsection{Metric potential}
One can note that $e^{\lambda(r=0)}=1$ and $e^{\nu(r=0)}=D^2$. So the metric components do not suffer from central singularities.
Variations in the metric potentials for compact star LMC X-4 are depicted in Fig.~\ref{metric}. It is clearly obvious from these graphical depictions that both metric potentials are free from geometrical and central singularities, have minimal values in the centre, increase nonlinearly and reach maximum values at the surface.
\subsection{Nature of pressure and density}
The central density and central pressure for our present model is obtained as,
\begin{eqnarray*}
\rho_c&=&\rho(r=0)=\frac{1}{4} (6 a \alpha + \beta),\\
p_c&=&p_r(r=0)=p_t(r=0)=-\frac{a \alpha}{2} + \alpha B - \frac{\beta}{4}.
\end{eqnarray*}

\begin{figure}[htbp]
    \centering
        \includegraphics[scale=.45]{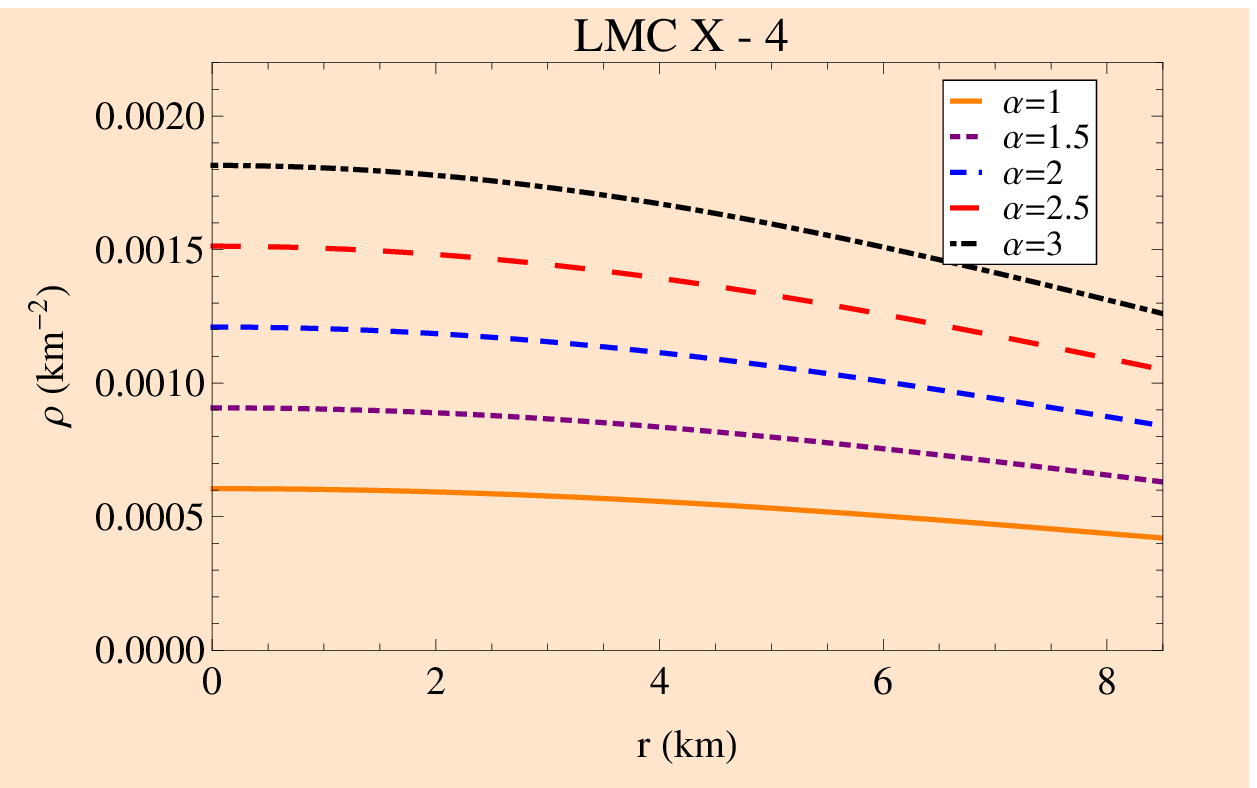}
        \includegraphics[scale=.45]{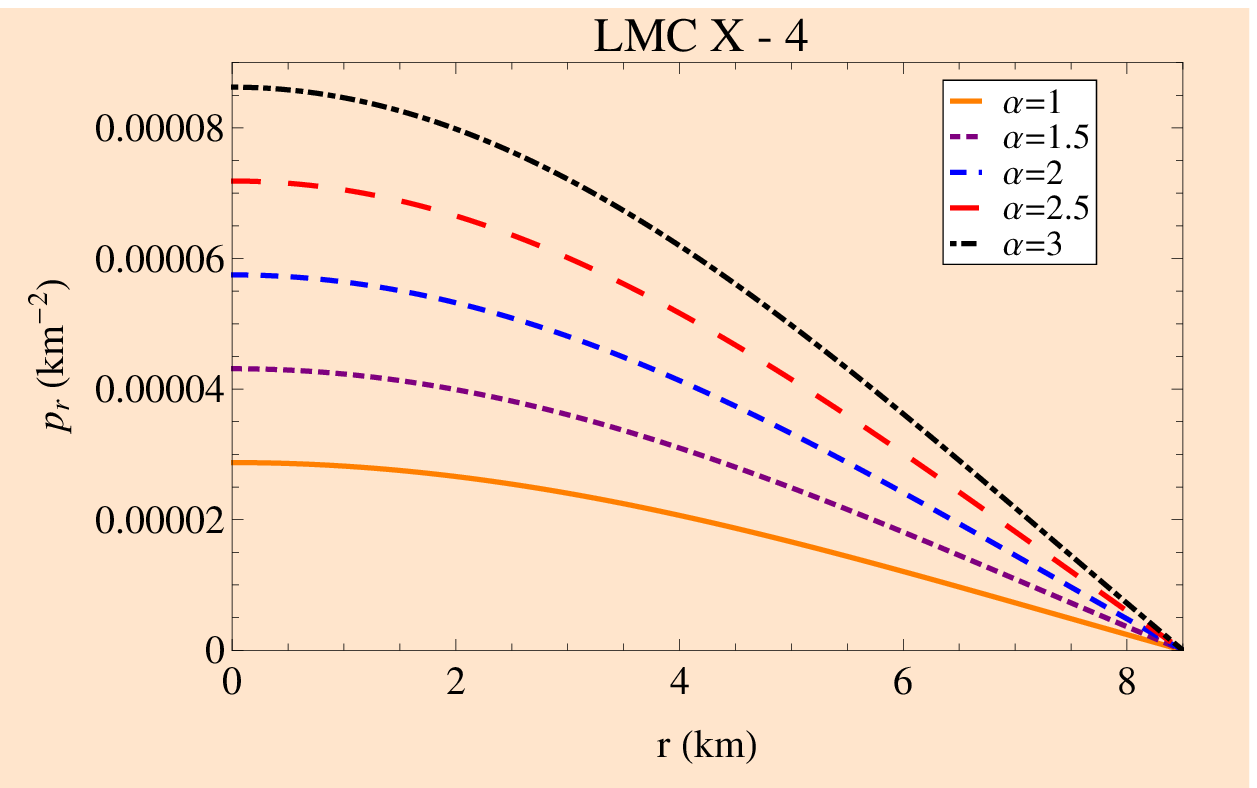}
        \includegraphics[scale=.45]{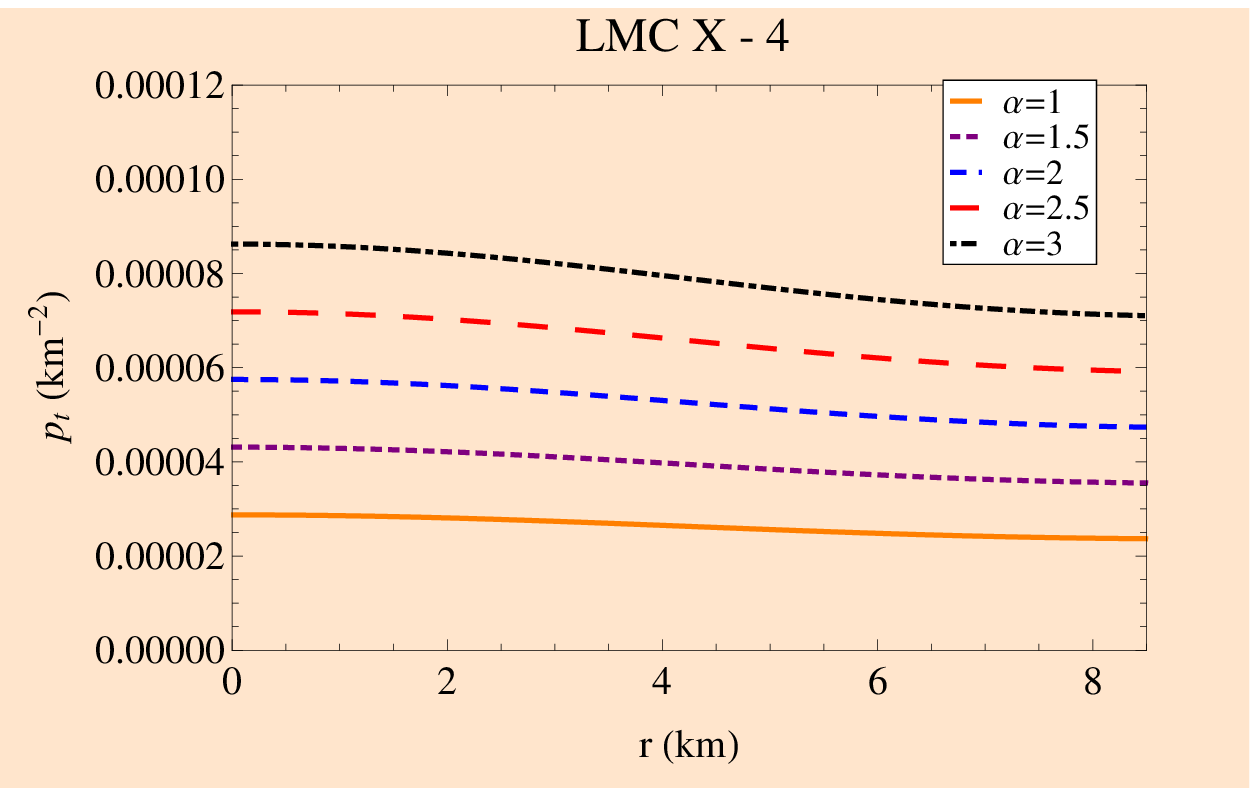}
       \caption{The matter density, radial and transverse pressure are shown for different values of $\alpha$}
    \label{rho1}
\end{figure}

Due to the densest nature of compact relativistic objects, the matter density and pressure components inside the stellar model exhibit their highest value at the center. In Fig. \ref{rho1}, the graphical analysis of the energy density, radial pressure and transverse pressure for compact star candidates are presented for different values of $\alpha$.

This graphical behavior makes it obvious that the matter density and pressure components reach their greatest values at the center of the compact stars and further approach to zero on the surface boundary, highlighting the fact that our stellar model exhibit very high compactness. For our recommended models under $f(T)$ gravity, these figures clearly show the existence of anisotropic structure of compact stars. Table~\ref{tbb2} display the numerical values of central density, surface density and central pressure for different values of $\alpha$. Our current system is free from physical and geometric singularities, which is supported by the fact that these physical properties are positive finite at their core.\\
Differentiating the expressions of density and both radial and transverse pressure, we obtain the following expressions:
\begin{eqnarray}
  \frac{d\rho}{dr}&=& -\frac{\alpha r \left(5 a^2 - 5 b + a (a^2 + 13 b) r^2 + 3 b (a^2 + 4 b) r^4 +
    3 a b^2 r^6 + b^3 r^8\right)}{4\pi(1 + a r^2 + b r^4)^3},\\
    \frac{dp_r}{dr}&=& \frac{\alpha r \left(a^2 - b - 2 a B + 2 b (a - 2 B) r^2 + b^2 r^4\right)}{4\pi(1 + a r^2 +
   b r^4)^2},\\
   \frac{dp_t}{dr}&=& \frac{\alpha r \left\{B^2 + a^2 (2 - B r^2) -
   b \big(2 + 8 B r^2 - 6 b r^4 + b B^2 r^8\big) +
   a \Big(6 b r^2 - B \big(3 + 3 b r^4 + B r^2 (-1 + b r^4)\big)\Big)\right\}}{4\pi(1 + a r^2 +
  b r^4)^3}.
   \end{eqnarray}

The gradients of density, radial and transverse pressure are represented by the symbols $\frac{d\rho}{dr},\,\frac{dp_r}{dr}$ and $\frac{dp_t}{dr}$ respectively.
\begin{figure}[htbp]
    \centering
        \includegraphics[scale=.45]{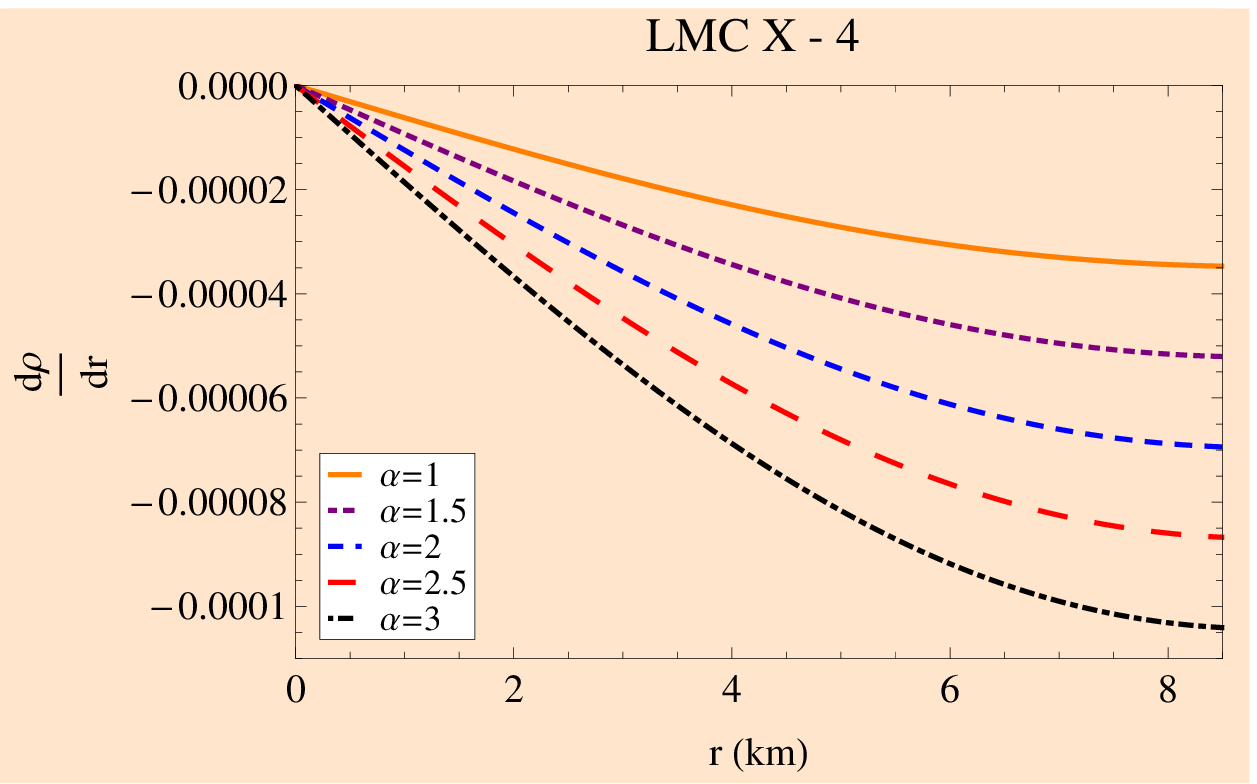}
        \includegraphics[scale=.45]{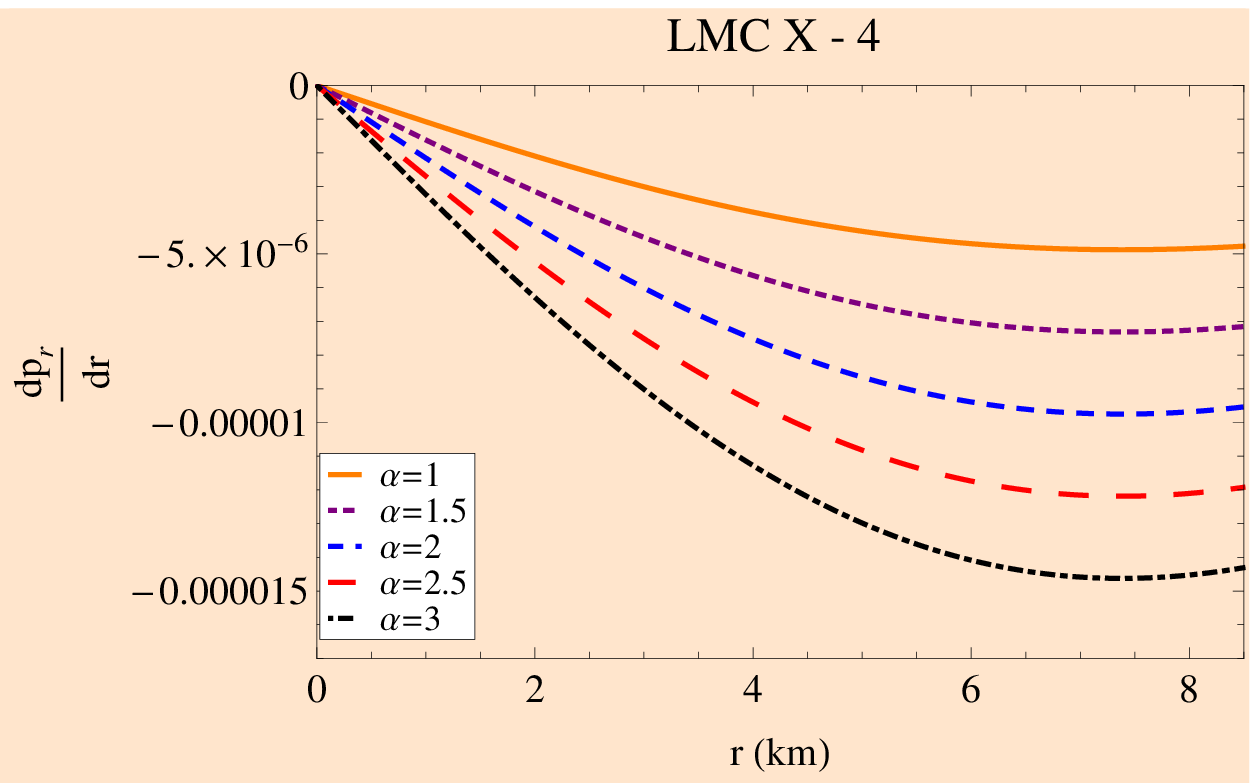}
        \includegraphics[scale=.45]{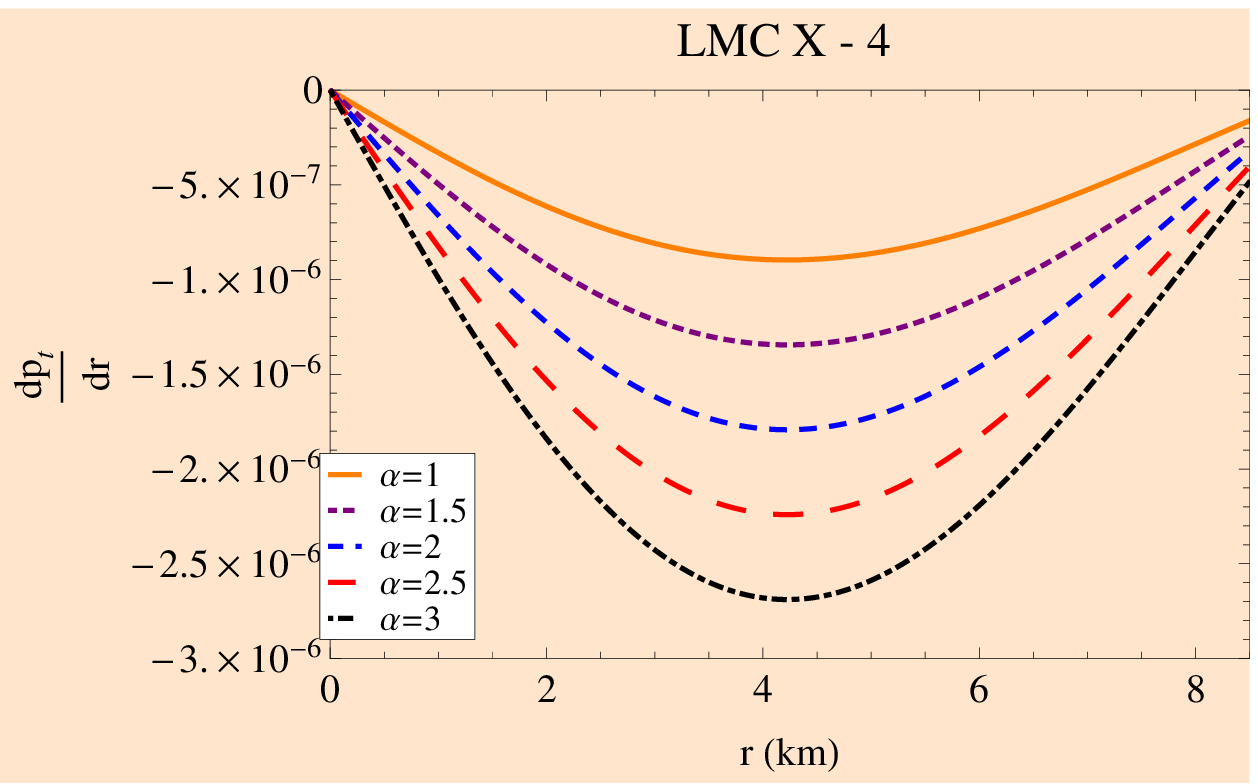}
       \caption{The density and pressure gradient are shown inside the stellar interior}
    \label{drho}
\end{figure}

Fig.~\ref{drho} provides a graphic illustration of these derivatives. From the figures it is clear that at the center of the star all the derivatives take zero value and $\rho',\,p_r',\,p_t'$ all take negative value in the interior of the star confirming that these quantities have maximum value at the center.
\subsection{Anisotropy}
\begin{figure}[htbp]
    \centering
        \includegraphics[scale=.45]{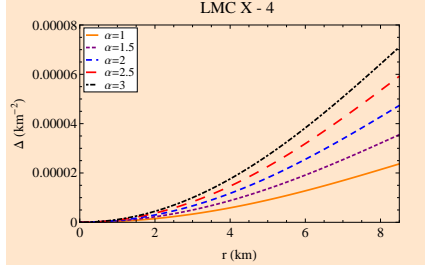}
       \caption{The anisotropic factor is shown for different values of $\alpha$}
    \label{delta}
\end{figure}

When modeling compact stars, the term anisotropy defined by $\Delta=p_t-p_r$ includes information on the anisotropic behavior of the model and it can be used to illustrate the internal structure of relativistic stellar objects. For our present model, the expression of anisotropy is given by
\begin{eqnarray}
\Delta &=& \frac{\alpha r^2 \left\{a^2 - a B + B^2 + a (2 b + B^2) r^2 -
   b (1+2 B r^2 - (b + B^2) r^4)\right\}}{8\pi (1 + a r^2 + b r^4)^2}.
   \end{eqnarray}
   Using observational data of the star LMC X-4 under consideration, we visually examine the anisotropy behavior of the star. If $p_t> p_r$ anisotropic pressure is directed outward, resulting in $\Delta>0$; however, if $p_t<p_r$, anisotropy turns negative, resulting in $\Delta<0$; this indicates that anisotropic pressure is being acted inward. As shown in Fig.~\ref{delta}, the graphical analysis of the anisotropic measurement with radial coordinate $r$ reveals that $p_t>p_r$ for different values of $\alpha$.
\subsection{Equation of state parameter}
\begin{figure}[htbp]
    \centering
        \includegraphics[scale=.45]{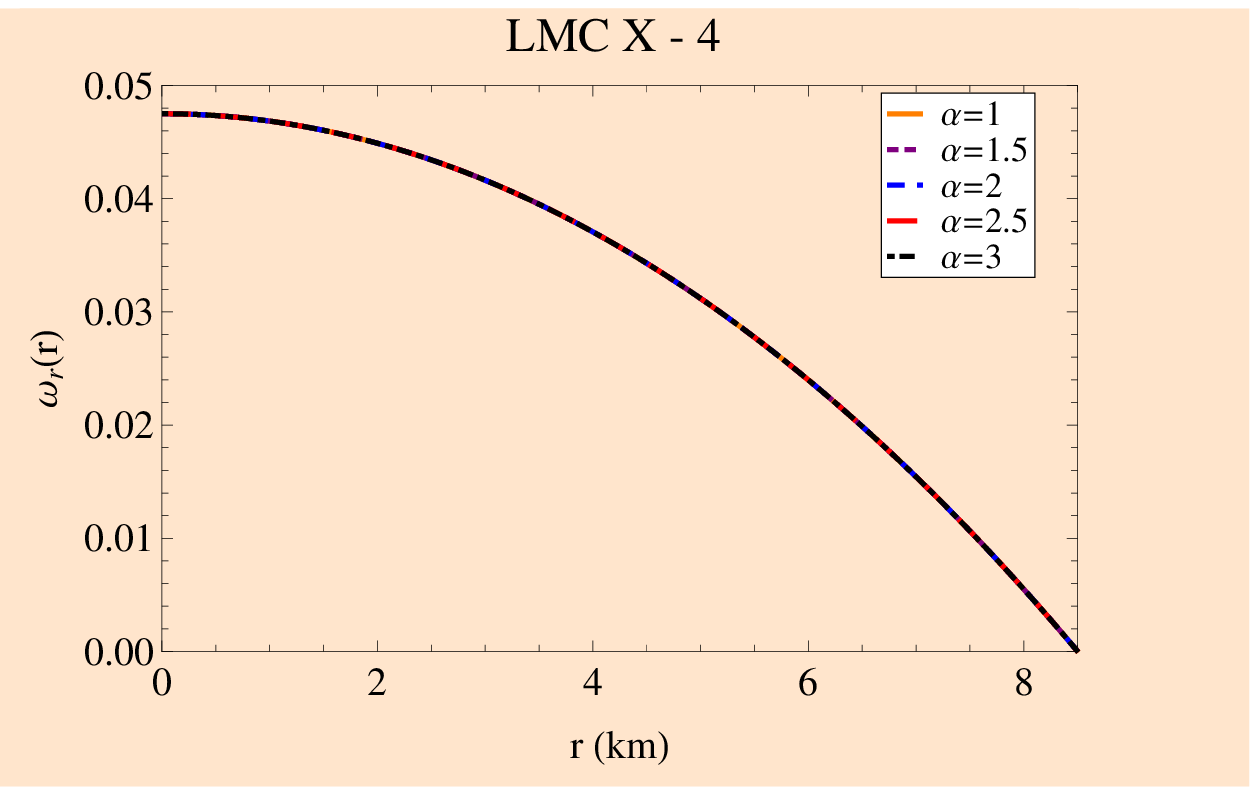}
        \includegraphics[scale=.45]{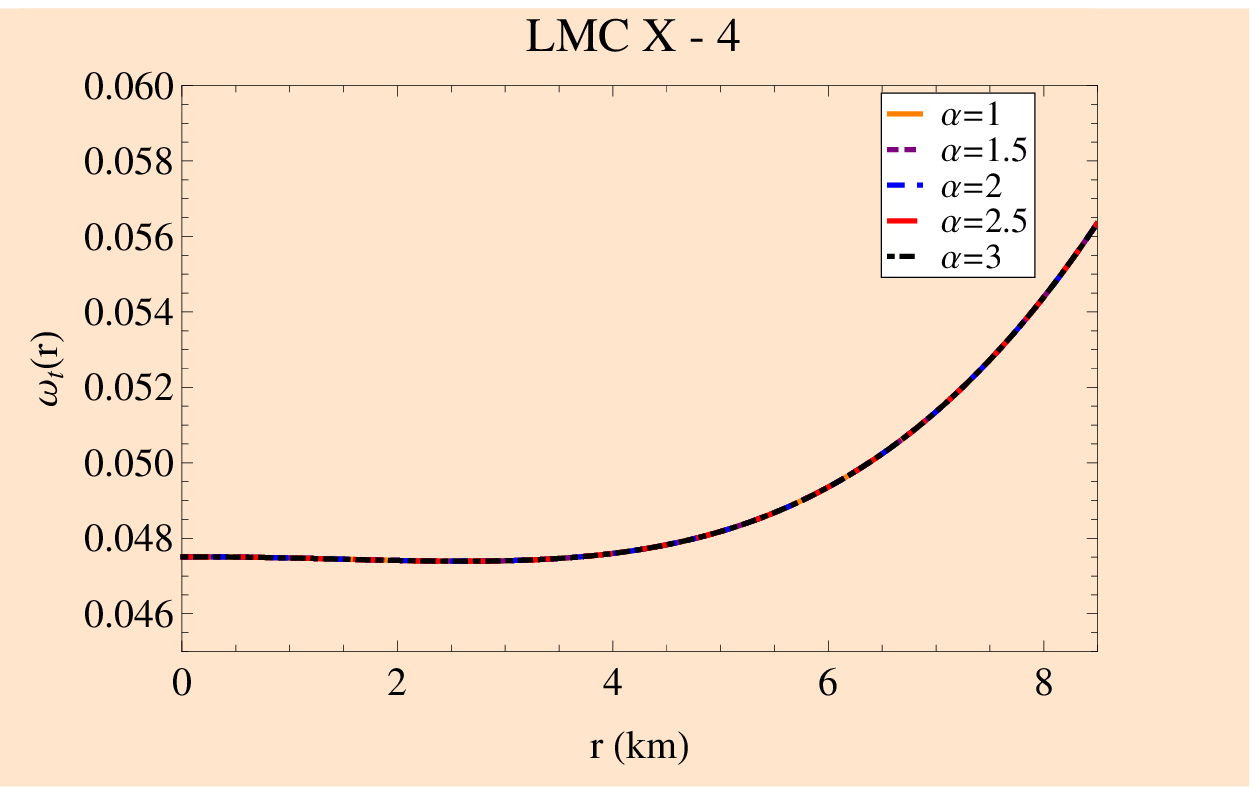}
       \caption{$\omega_r$ and $\omega_t$ are shown inside the stellar interior}
    \label{omegar}
\end{figure}

\begin{figure}[htbp]
    \centering
        \includegraphics[scale=.45]{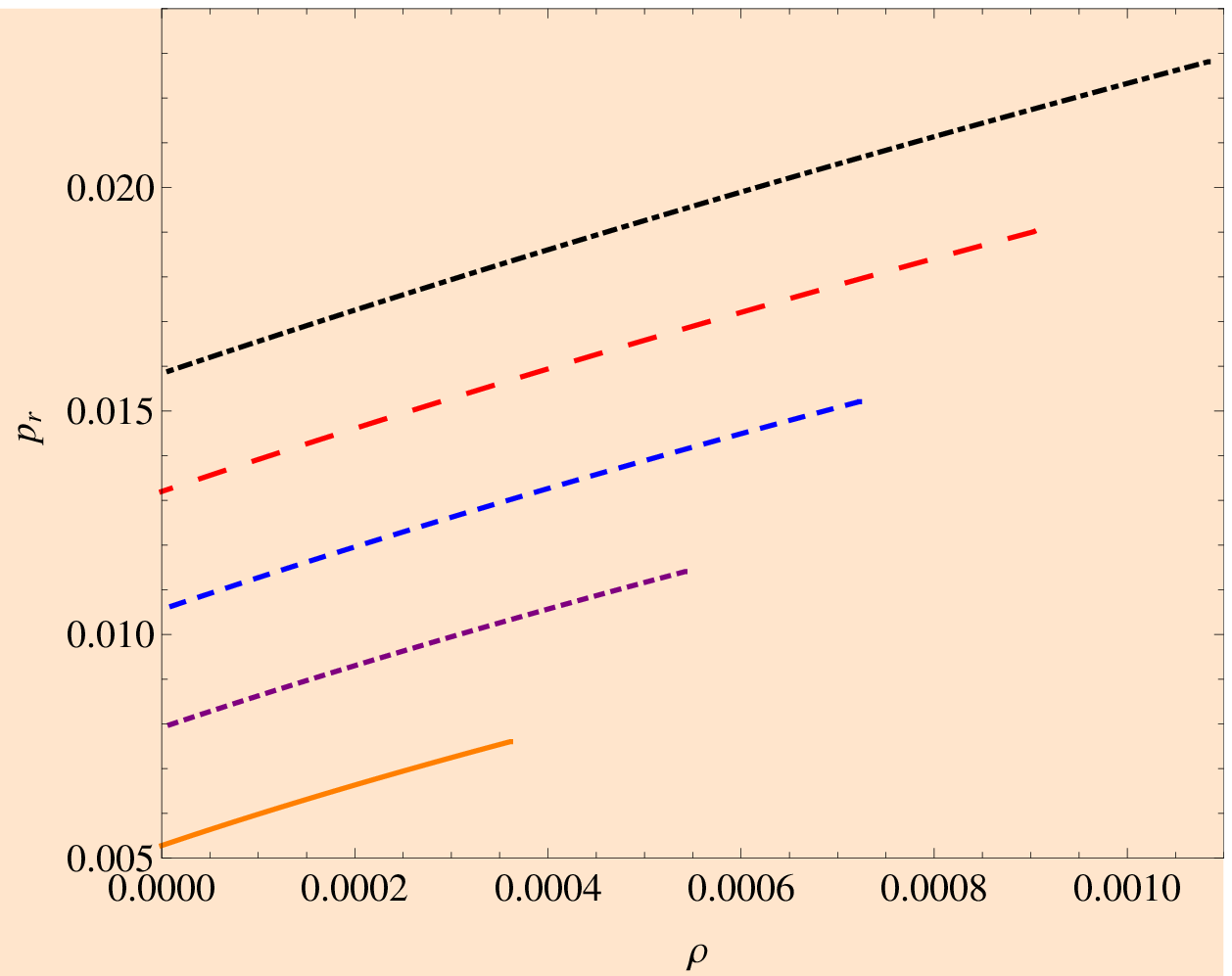}
        \includegraphics[scale=.45]{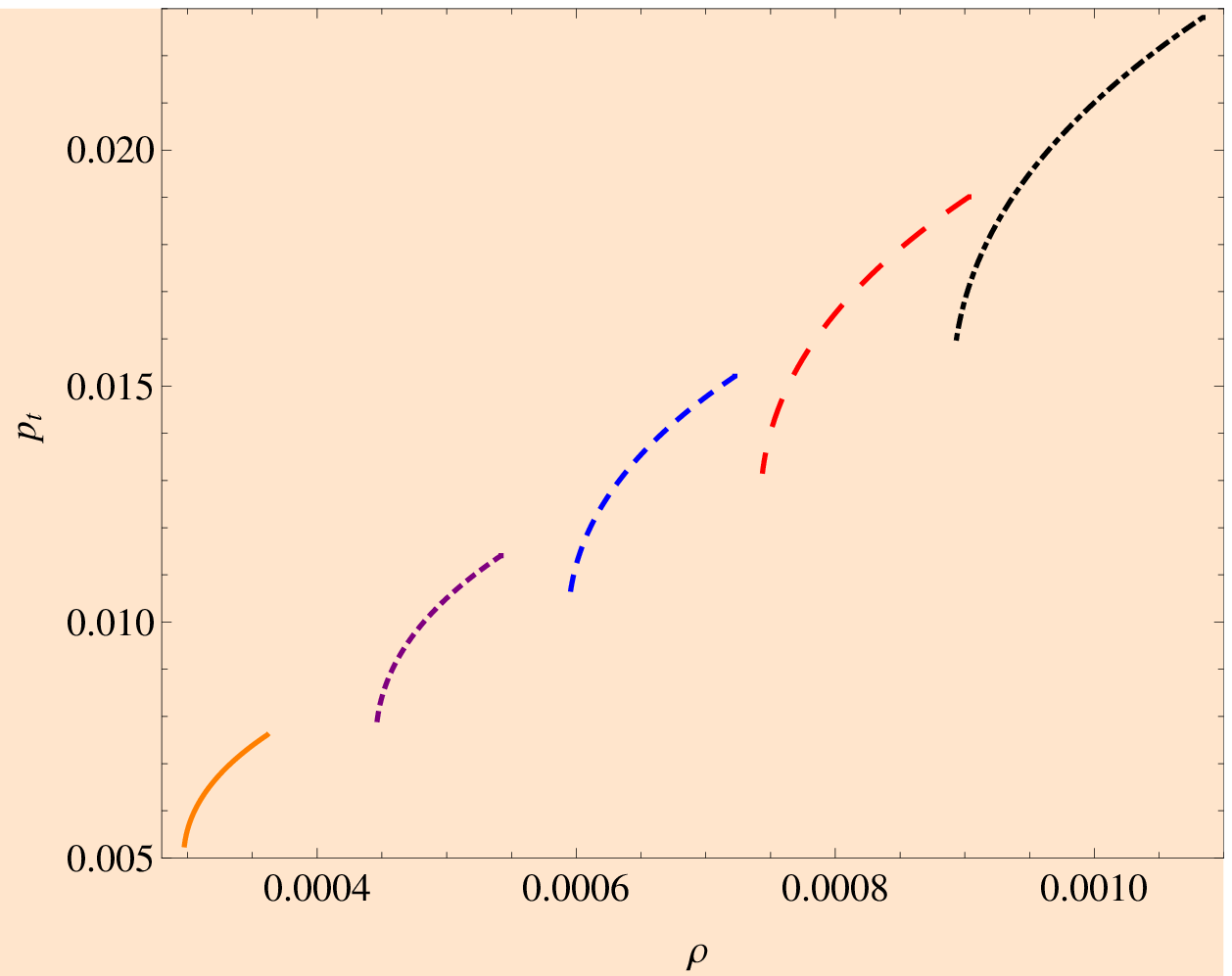}
       \caption{The variation of pressure with respect to density are shown inside the stellar interior}
    \label{e11}
\end{figure}

The equation of state parameters $\omega_r$ and $\omega_t$ for our present model are obtained as,
\begin{eqnarray*}
\omega_r=\frac{p_r}{\rho}= \frac{(1 + a r^2 + b r^4) (4 \alpha B-2 a \alpha
   -\beta - (2 \alpha b + a \beta) r^2 - b \beta r^4)}{6 a \alpha + \beta +
 2 (a^2 \alpha + 5 \alpha b + a \beta) r^2 + (4 a \alpha b + a^2 \beta +
    2 b \beta) r^4 + 2 b (\alpha b + a \beta) r^6 + b^2 \beta r^8} ,\\
\omega_t=\frac{p_t}{\rho}= \frac{2 \alpha (2 B-a) - \beta -
 2 \Big(\alpha (2 b - B (a + B)) + a \beta\Big) r^2 + (2 a \alpha B^2 -
    a^2 \beta - 2 b \beta) r^4 + 2 b (\alpha B^2 - a \beta) r^6 -
 b^2 \beta r^8}{6 a \alpha + \beta +
 2 (a^2 \alpha + 5 \alpha b + a \beta) r^2 + (4 a \alpha b + a^2 \beta +
    2 b \beta) r^4 + 2 b (\alpha b + a \beta) r^6 + b^2 beta r^8}.
\end{eqnarray*}
The profiles of these two parameters are shown in Fig.~\ref{omegar} for different values of $\alpha$. One can note that all plots coincide for different values of $\alpha$. $\omega_r$ is monotonic decreasing function of $r$, whereas $\omega_t$ is monotonic increasing function of $r$. Both $\omega_r$ and $\omega_t$ lie in the range $0<\omega_r,\,\omega_t<1$.\\
For develop the compact star model in this paper we have not assumed any particular equation of state. In Fig.~\ref{e11}, we have shown the variation of pressures with respect to density for different values of $\alpha$. From the figure it is clear that the radial pressure varies linearly with respect to density but the transverse pressure varies non linearly with respect to $\rho$.

\subsection{Sound Velocity}

For a physically stable stellar structure, the radial and transverse sound speed must be in the range (0,\,1), i.e., $0<V_r^2<1$ and $0<V_t^2<1$, everywhere inside the star. To investigate the possible stable/unstable configurations of stellar structures, Herrera and coworkers \cite{Herrera:1992lwz,10.1093/mnras/265.3.533,DiPrisco:1997tw} developed a new notion of the cracking concept. The difference in sound propagation can be used to determine potentially stable/unstable regions within matter configurations.
Potentially stable regions are those where radial sound speed components are greater than transverse sound speed components, i.e., $V_r^2-V_t^2>0$, whereas unstable regions do not exhibit this inequality.
For our present model, the square of radial and transverse speed of sound are calculated as,
\begin{eqnarray}
V_r^2=\frac{dp_r}{d\rho}&=&\frac{(1 + a r^2 + b r^4) \Big(b-a^2+2 a B - 2 b (a - 2 B) r^2 -
   b^2 r^4\Big)}{
5 a^2 - 5 b + a (a^2 + 13 b) r^2 + 3 b (a^2 + 4 b) r^4 + 3 a b^2 r^6 +
  b^3 r^8},\\
V_t^2=\frac{dp_t}{d\rho}&=&\frac{-B^2 - a^2 (2- B r^2) + b (2 + 8 B r^2 - 6 b r^4 + b B^2 r^8) -
 a \left\{6 b r^2 -B \Big(3 + 3 b r^4 + B r^2 (-1 + b r^4)\Big)\right\}}{
5 a^2 - 5 b + a (a^2 + 13 b) r^2 + 3 b (a^2 + 4 b) r^4 + 3 a b^2 r^6 +
  b^3 r^8}
\end{eqnarray}

\begin{figure}[htbp]
    \centering
        \includegraphics[scale=.45]{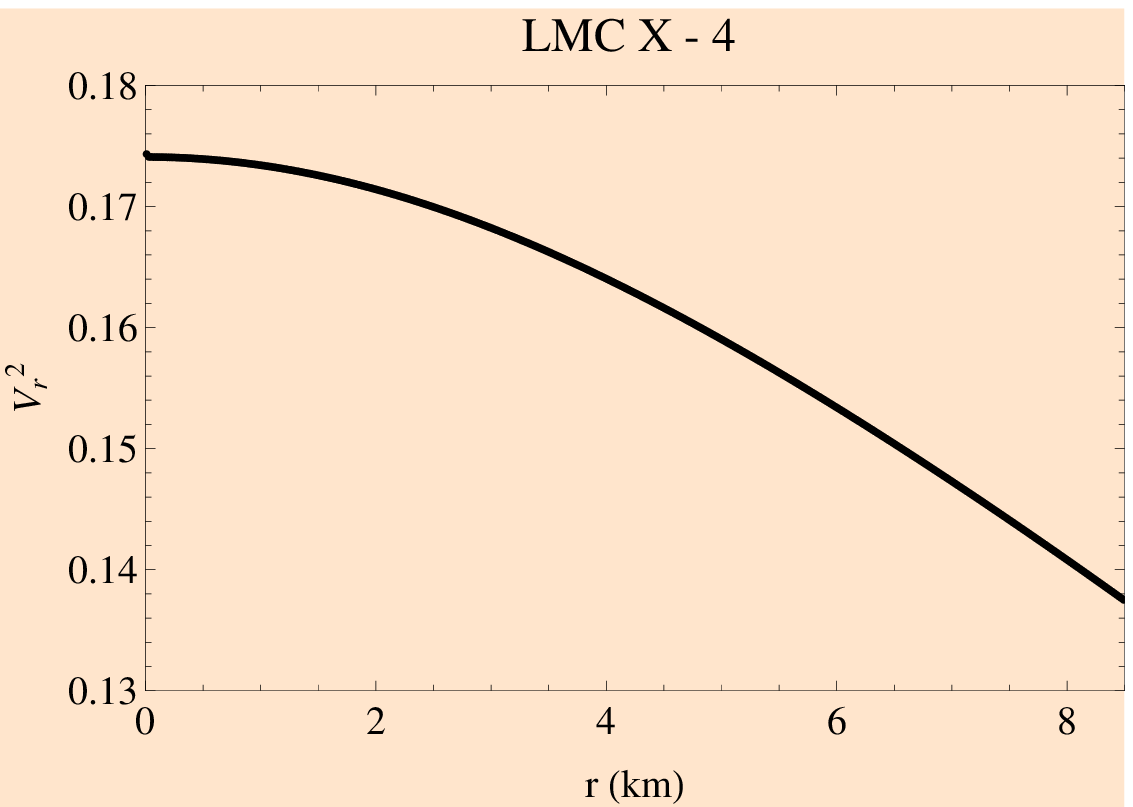}
        \includegraphics[scale=.45]{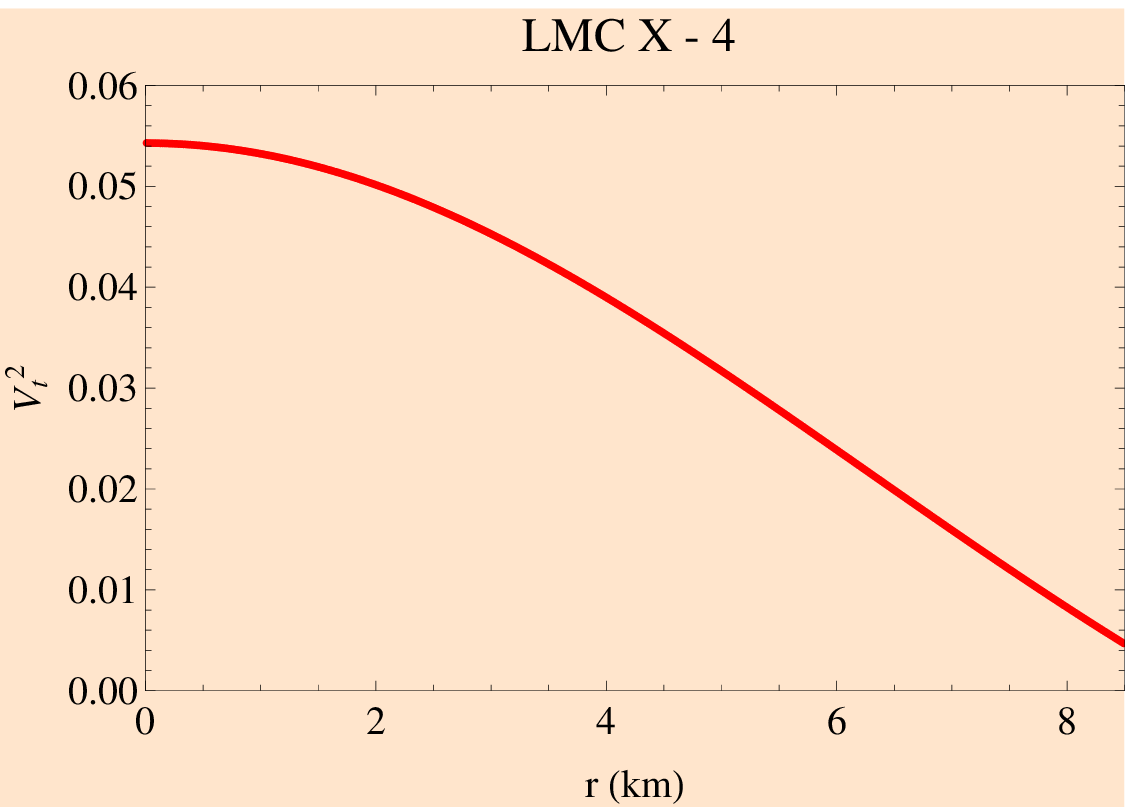}
        \includegraphics[scale=.45]{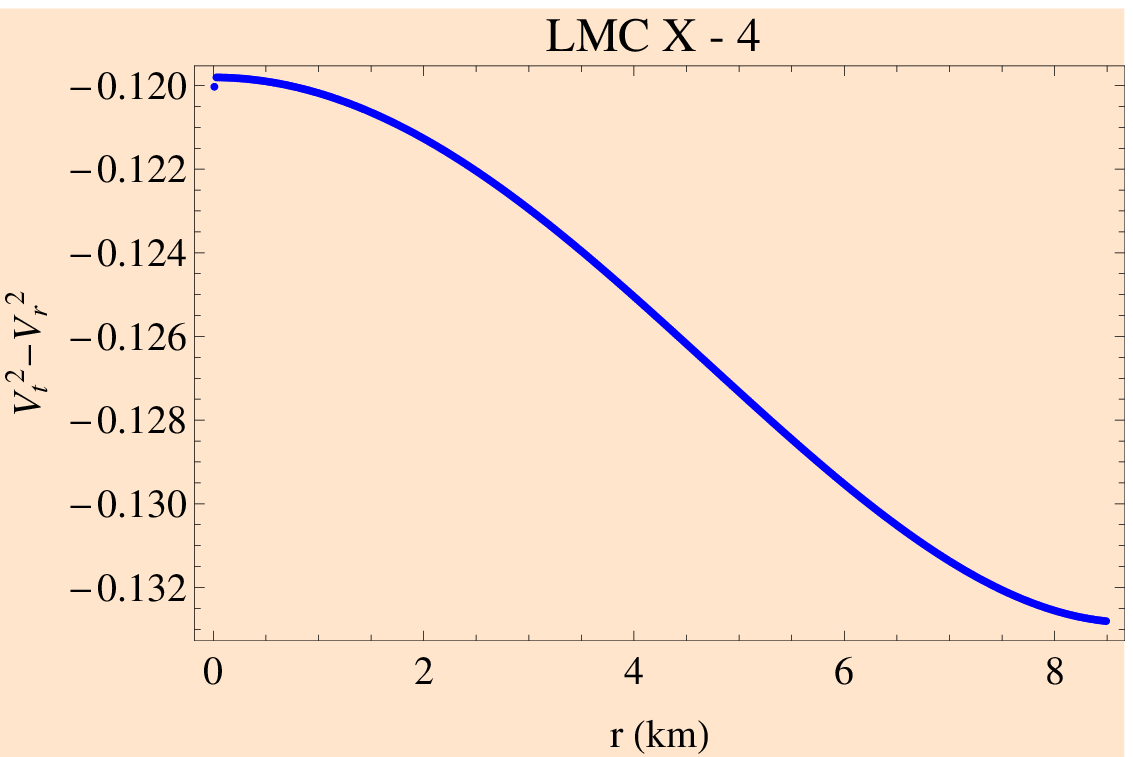}
       \caption{The square of the sound velocity and stability factor $V_t^2-V_r^2$ is shown against $r$}
    \label{svt}
\end{figure}

From Fig~\ref{svt}, it is seen that $0<V_r^2,\,V_t^2<1$ and the figure also show that how the radial and transverse speeds of sounds behaved for the compact star candidate LMC X 4. It is also noticed that the matter configuration relation is stable as aforementioned since $V_r^2>V_t^2$ everywhere inside the star. It is also interesting to note that the both the sound speed does not depend on $\alpha$.

\subsection{Adiabatic index}
\begin{figure}[htbp]
    \centering
        \includegraphics[scale=.45]{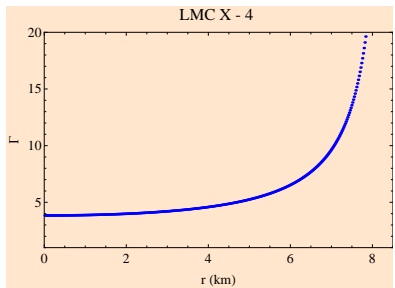}
       \caption{Relativistic adiabatic index is shown inside the stellar interior}
    \label{gammar}
\end{figure}

For a given energy density, the adiabatic index can be used to explain the stiffness of the equation of state. This term also serves to illustrate the stability of both relativistic and non-relativistic compact stars. Chandrasekhar \cite{Chandrasekhar:1964zz} (as a pioneer) proposed the idea of dynamical stability against infinitesimal radial adiabatic perturbation of the stellar system. Subsequently, this concept was successfully proven by several authors \cite{1975A&A....38...51H,hillebrandt1976anisotropic,Horvat:2010xf,Doneva:2012rd,Silva:2014fca,bombaci1996maximum} for both isotropic and anisotropic stellar objects. According to their study, the adiabatic index for dynamically stable stellar objects must be greater than $4/3$ at all internal points. $\Gamma$ is the notation of the adiabatic index and its expression for an anisotropic fluid is given by,
\begin{eqnarray}
\Gamma&=&\frac{\rho+p_r}{p_r}V_r^2,\nonumber\\
&=&\frac{4 \alpha \Big(a + B + (2 b + a B) r^2 + b B r^4\Big)}{(1 + a r^2 + b r^4) (4 \alpha B-2 a \alpha-
   \beta - (2 \alpha b + a \beta) r^2 - b \beta r^4)}V_r^2.
\end{eqnarray}
The profile of $\Gamma$ is shown in Fig.~\ref{gammar}. It can be noted from the figure that though $\alpha$ presents in the expression of $\Gamma$, the profiles of the adiabatic index coincides for different values of $\alpha$ and hence the expression of the adiabatic index does not depend on $\alpha$. Moreover the profile is monotonic increasing as well as greater than $4/3$ everywhere inside the stellar interior.

  \subsection{Mass radius relation and surface redshift}
\begin{figure}[htbp]
    \centering
        \includegraphics[scale=.45]{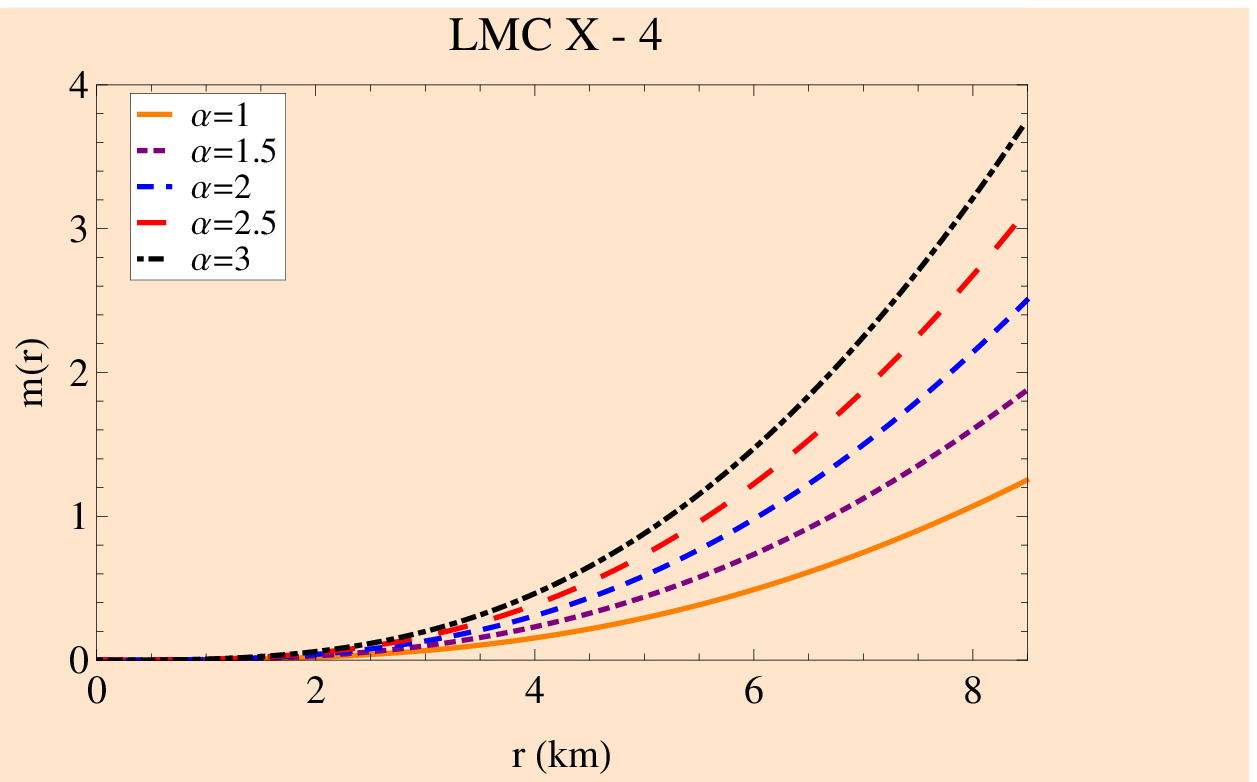}
        \includegraphics[scale=.45]{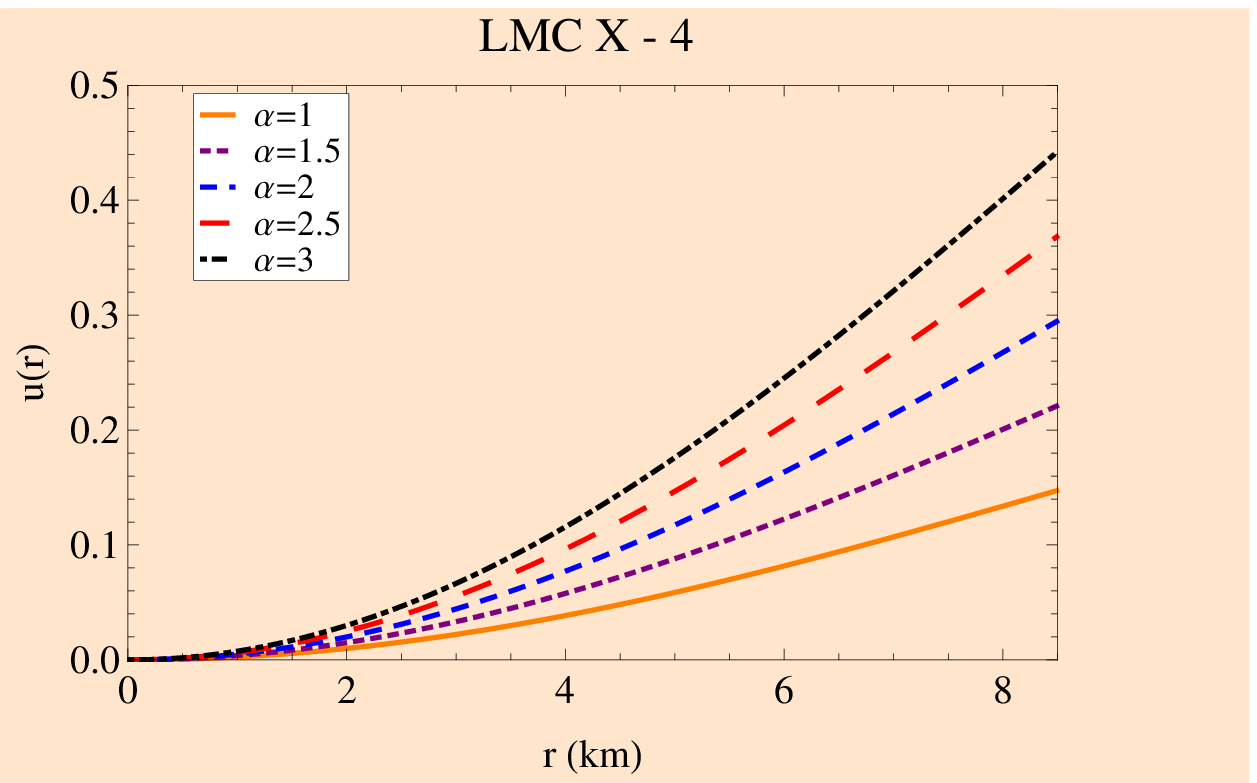}
        \includegraphics[scale=.45]{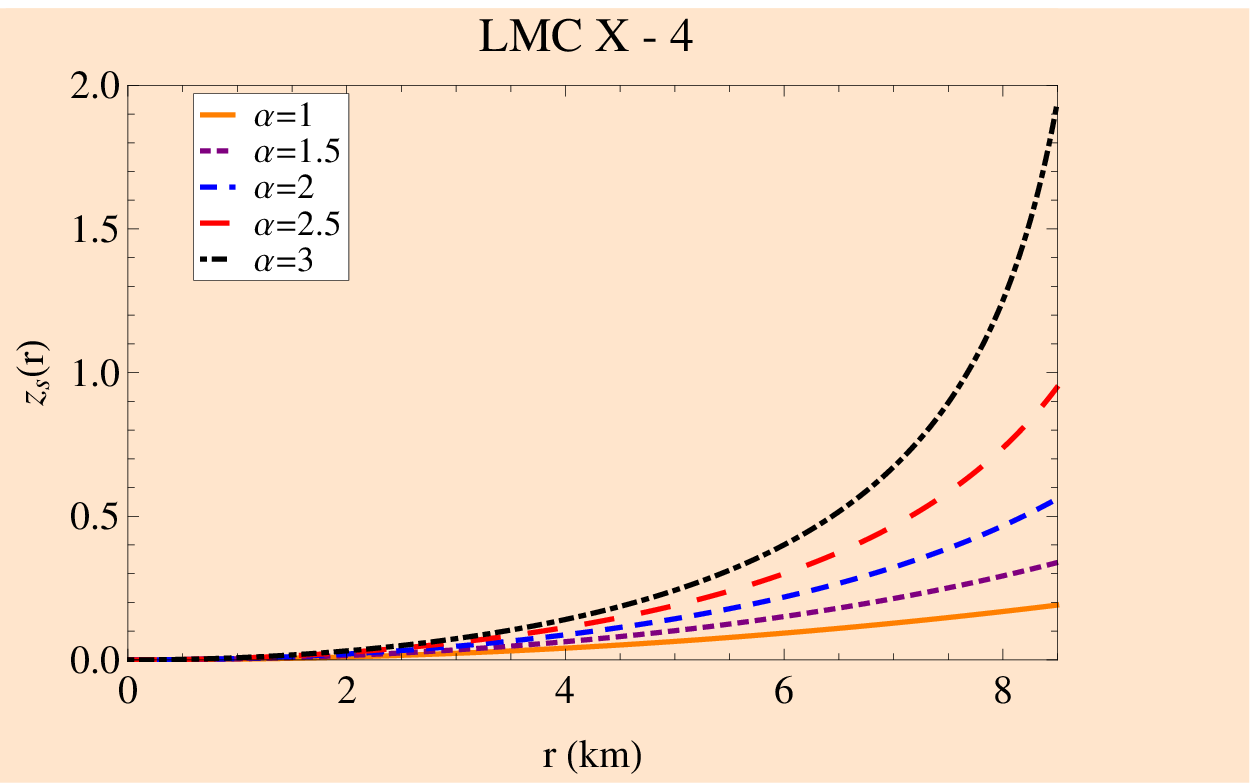}
       \caption{The mass function, compactness and surface redshift are shown for different values of $\alpha$}
    \label{mass1}
\end{figure}

  The mass function of our present model is calculated as,
  \begin{eqnarray}
  m(r)=\int_0^r 4\pi \rho r^2 dr=\frac{r^3}{12}\left[\frac{6\alpha(a+br^2)}{1+ar^2+br^4}+\beta\right],
  \end{eqnarray}
  the compactness $u(r)$ is obtained from the formula $u(r)=\frac{m(r)}{r}=\frac{r^2}{12}\left[\frac{6\alpha(a+br^2)}{1+ar^2+br^4}+\beta\right]$.\\
  The surface redshift $z_s$ can be obtained as,
  \begin{eqnarray}
  z_s=\left(1-2\frac{m(r)}{r}\right)^{-\frac{1}{2}}-1
  \end{eqnarray}
  The profiles of mass function, compactness and surface redshift are shown in fig.~\ref{mass1} for different values of $\alpha$ mentioned in the figure. All functions are monotonic increasing functions of `r'. Surface redshift can be used to explain the strong physical interaction between the internal particles of the star and its equation of state. In our context, compact star satisfy the Buchdahl requirement $u(R)<\frac{4}{9}$ \cite{Buchdahl:1959zz}.
  Barraco and Hamity \cite{Barraco:2002ds} demonstrated that, when the Cosmological constant is absent, the redshift $z_s$ for an isotropic star must be $\leq 2$. After that, Bohmer and Harko \cite{Boehmer:2006ye} generalised the aforementioned result for the case of anisotropy along with the Cosmological constant and discovered that $z_s \leq 5$. But Ivanov \cite{Ivanov:2002xf} shown that modifications or restriction leads the highest allowable value $ \leq~5.211$ in one of his pioneering publications. In Table~\ref{tbb2} we have calculated the values of surface redshift for different values of $\alpha$, meaning that our model predicts a stable compact star.

\subsection{Energy condition}
The null energy condition (NEC), weak energy condition (WEC), strong energy condition (SEC) and dominant energy condition (DEC) must all be satisfied inside the interior points of the stellar configuration to be considered physically valid. All the energy conditions are well satisfied if and only if the following inequalities hold good inside the stellar interior \cite{Biswas:2019doe} :\\
NEC : $\rho \geq 0$, \\
WEC : $\rho + p_r \geq 0$, $\rho + p_t \geq 0$, \\
SEC : $\rho + p_r + 2 p_t \geq 0$, \\
DEC : $\rho- |p_r| \geq 0$,~~$\rho-|p_t| \geq 0$.

\begin{figure}[htbp]
    \centering
        \includegraphics[scale=.45]{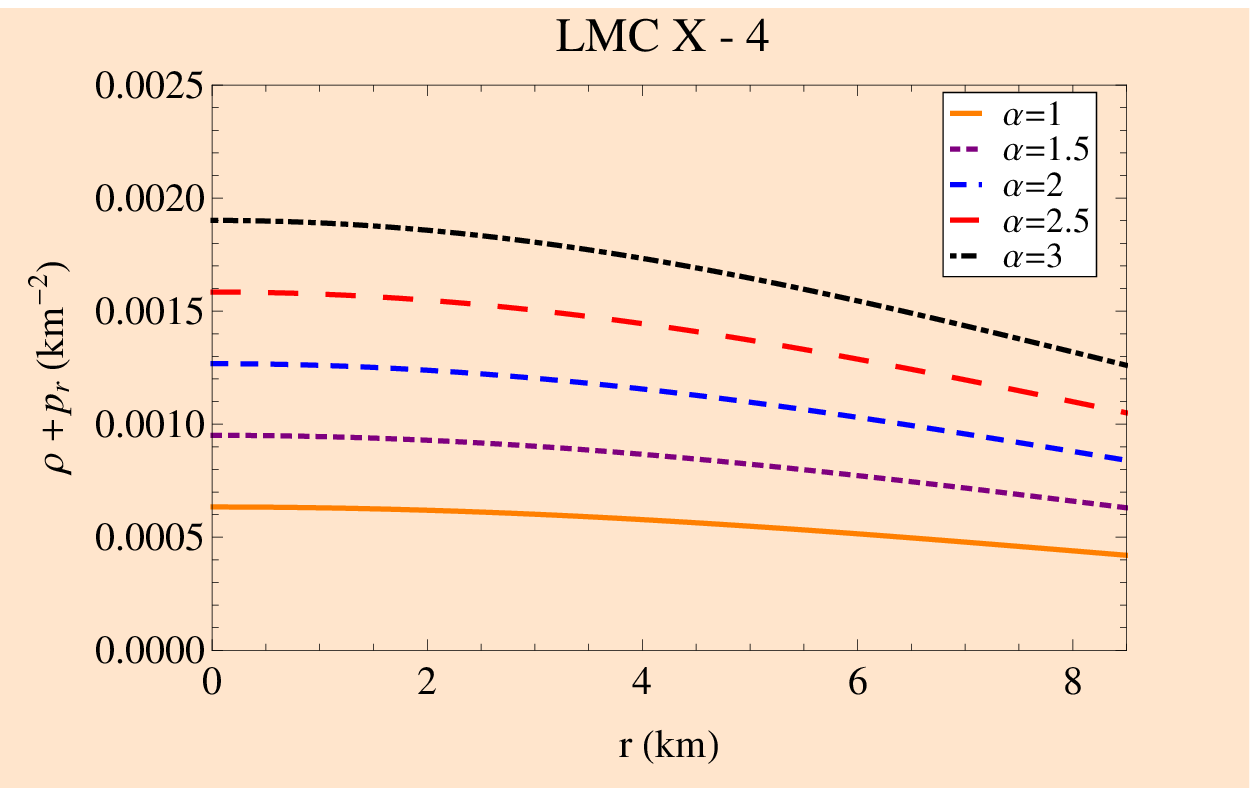}
        \includegraphics[scale=.45]{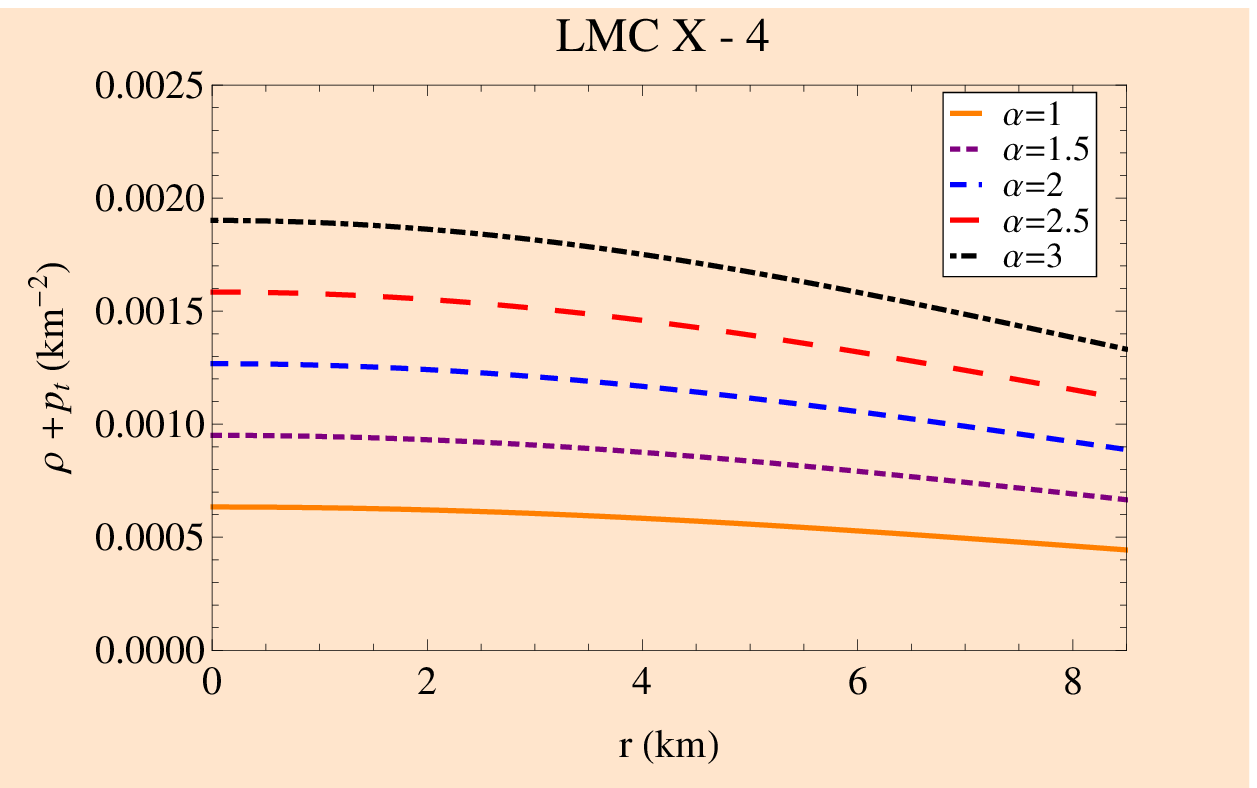}
        \includegraphics[scale=.45]{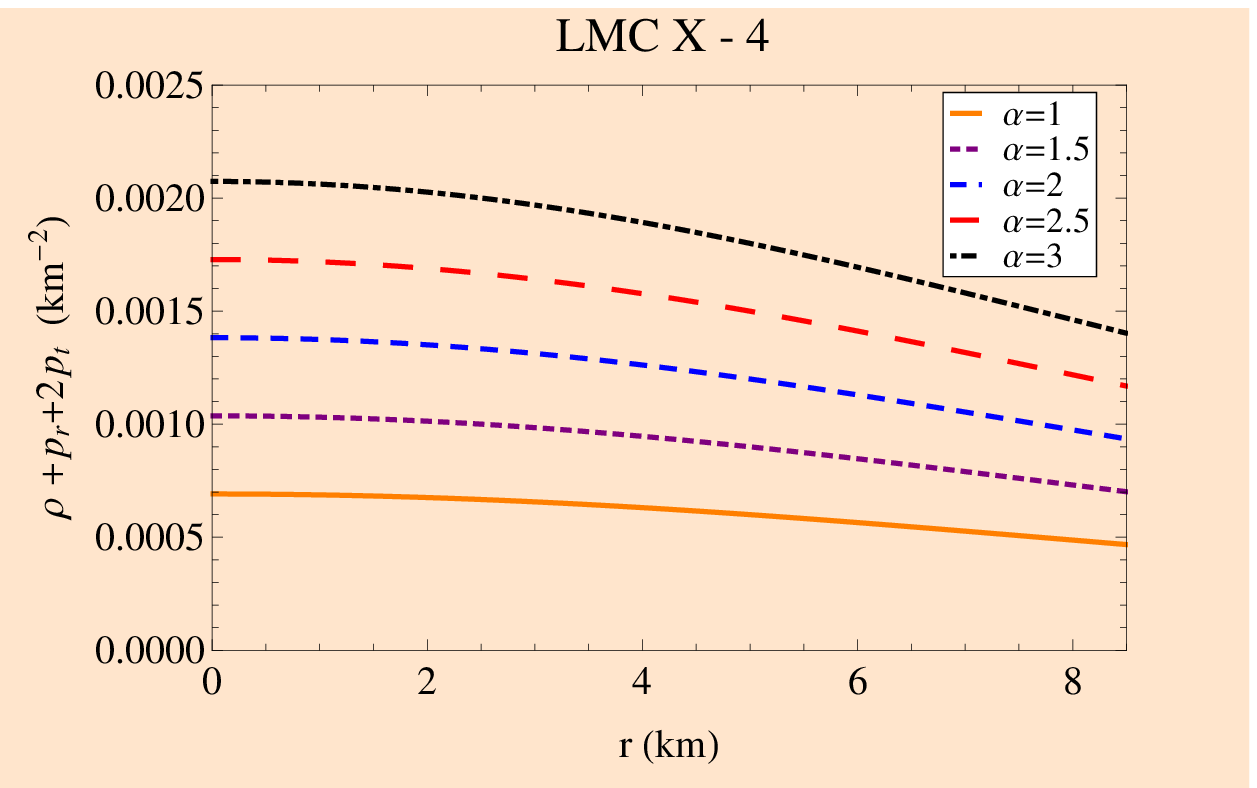}
        \includegraphics[scale=.45]{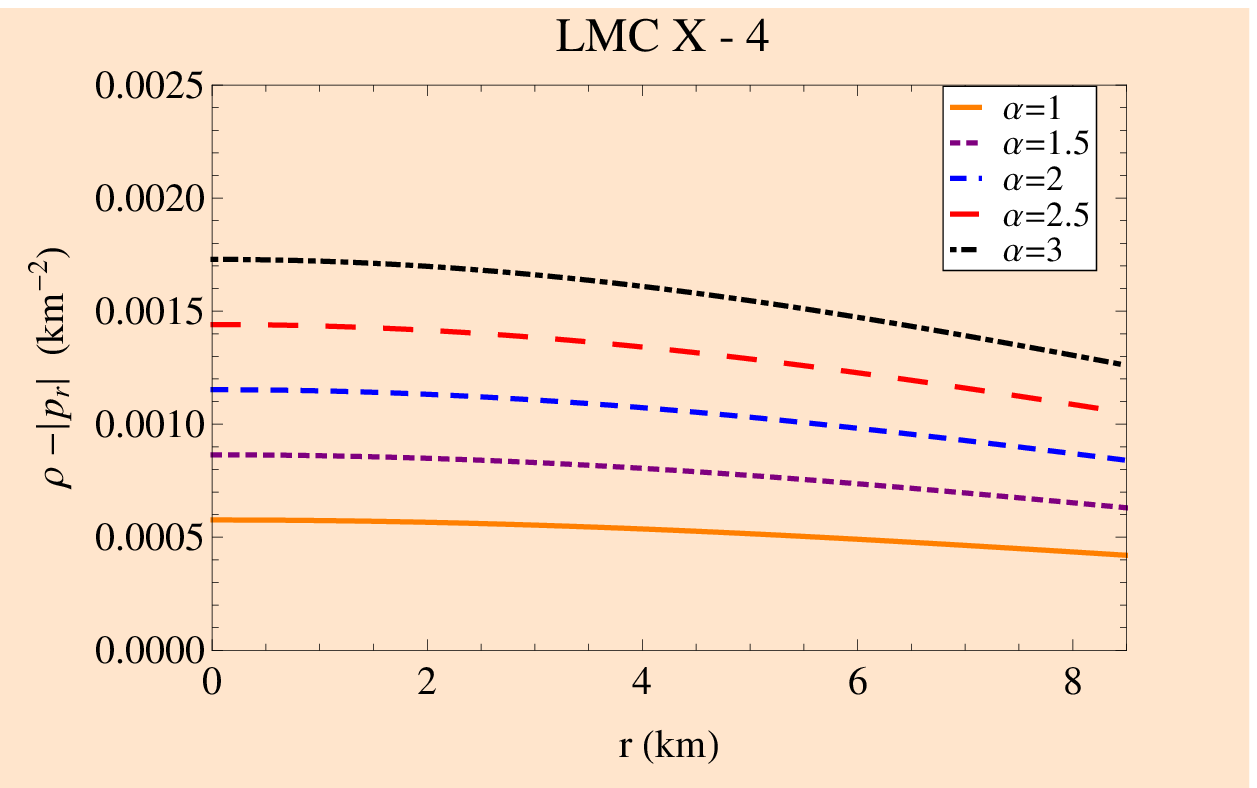}
        \includegraphics[scale=.45]{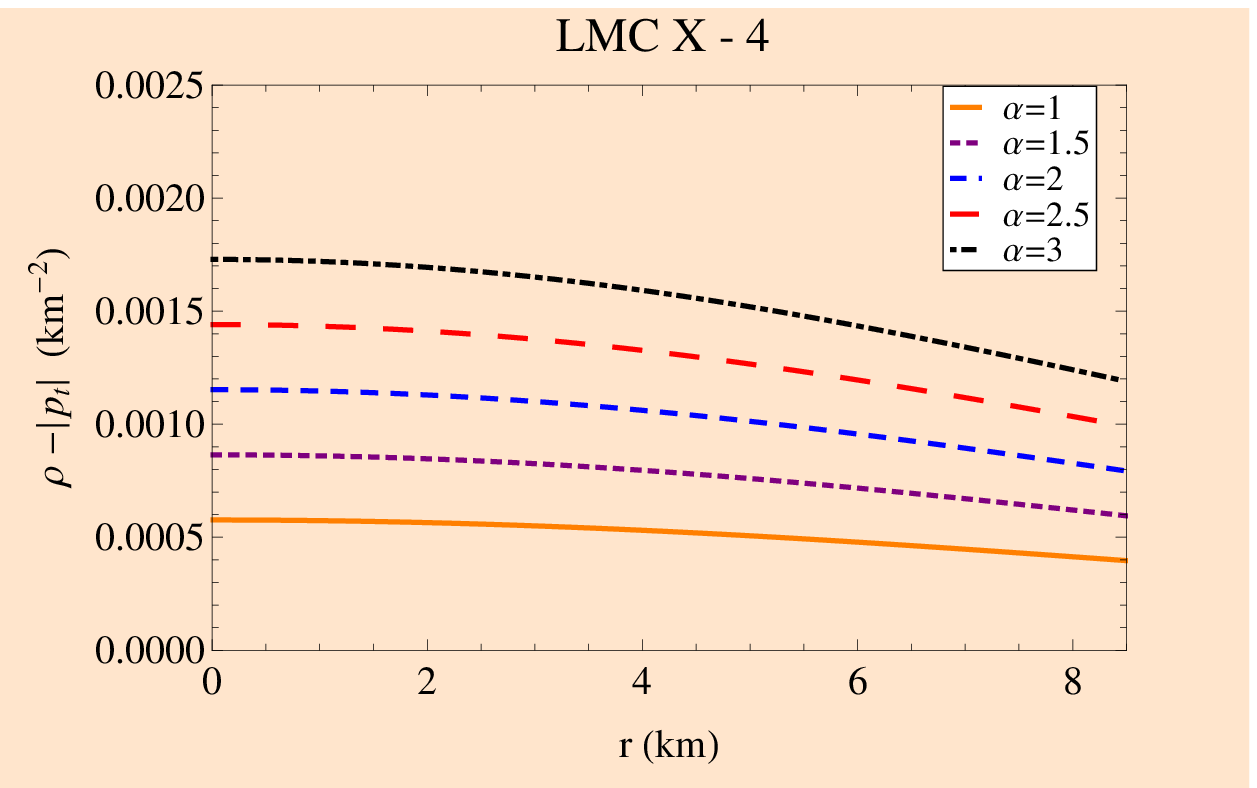}
       \caption{The energy conditions are shown against `r'}
    \label{ec2}
\end{figure}

For the compact star candidate LMC X-4, we have depicted the energy conditions. The graphical variation in Fig.~\ref{ec2} shows that the energy conditions are met and our model is very consistent with them.

\subsection{ Harrison-
Zeldovich-Novikov static stability criterion }

Chandrasekhar \cite{Chandrasekhar:1964zza}, Harrison et al. \cite{harrison1965gravitation} and others calculated the eigen-frequencies for each of the fundamental modes to evaluate the stability of stars. Later, Zeldovich and Novikov \cite{1971reas.book.....Z} simplified the calculations in accordance to Harrison et al. \cite{harrison1965gravitation}. From their analysis, they show that the model is stable if $\frac{\partial M}{\partial \rho_c} >0$ and for unstable model, the inequality will be in reverse direction. Where $\rho_c$ is the central density and the expression of $M$ is given as,

\begin{eqnarray}\label{t1}
M(\rho_c)=\frac{R^3 \left[24 \alpha \rho_c + 36 \alpha^2 b R^2 + \beta (-\beta + 4 \rho_c) R^2 +
   6 \alpha b \beta R^4\right]}{12 \left[6 \alpha + (-\beta + 4 \rho_c) R^2 + 6 \alpha b R^4\right]}.
   \end{eqnarray}
Now differentiating the Eqn. (\ref{t1}) partially we get,
\begin{eqnarray}
\frac{\partial M}{\partial \rho_c} &=& \frac{12 \alpha^2 R^3}{\left[(\beta - 4 \rho_c) R^2 - 6 \alpha (1 + b R^4)\right]^2}.
\end{eqnarray}

\begin{figure}[htbp]
    \centering
    \includegraphics[scale=.45]{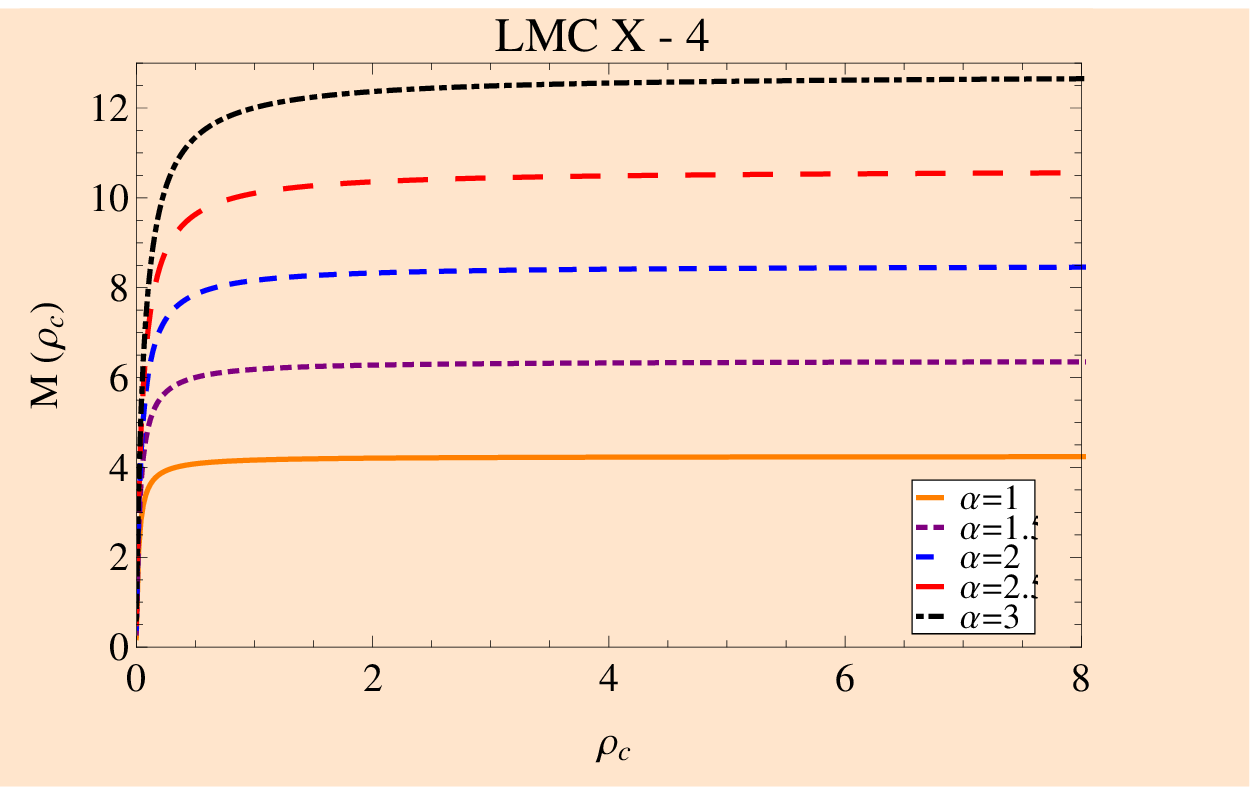}
        \includegraphics[scale=.45]{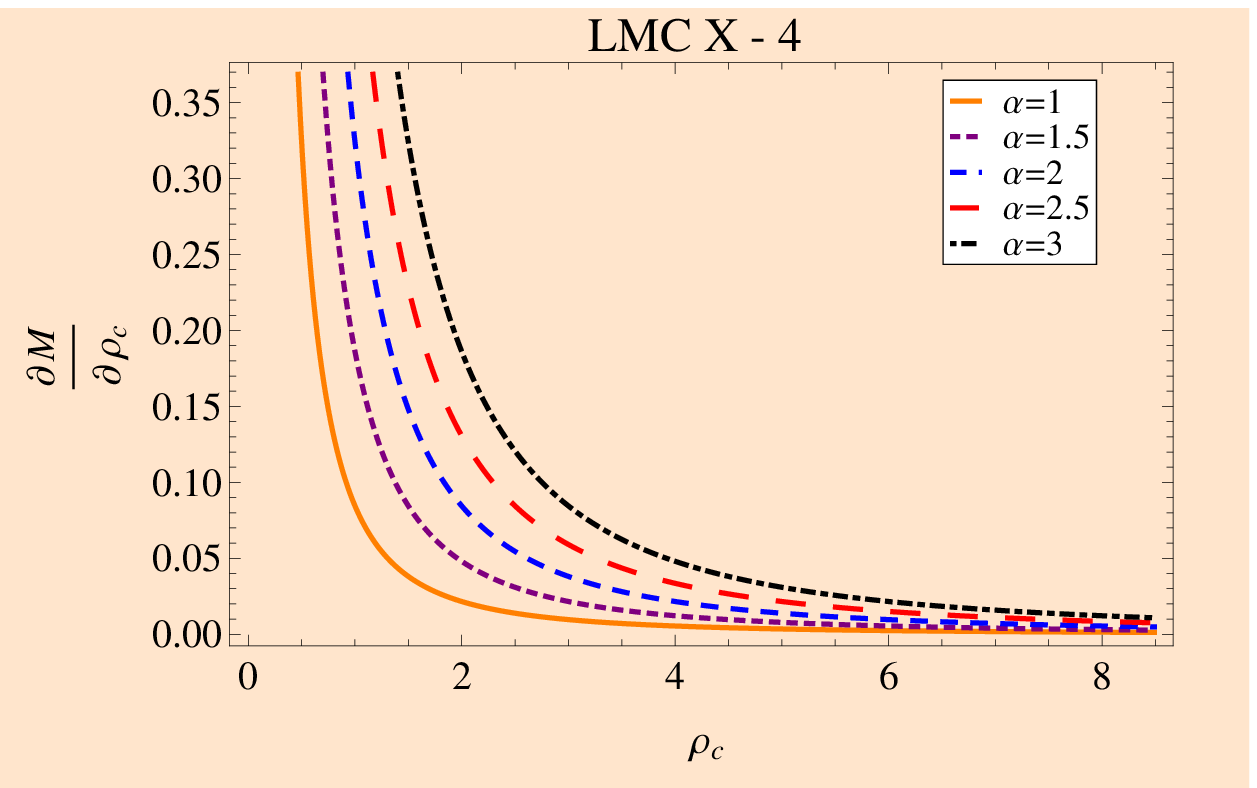}
       \caption{(left) $M(\rho_c)$ is shown against `$\rho$' and (right) $\frac{\partial M}{\partial \rho_c}$ is shown against `$\rho_c$' for different values of $\alpha$}
    \label{mass2}
\end{figure}

The profile of $M(\rho_c)$ versus $\rho_c$ and $\frac{\partial M}{\partial \rho_c}$ versus $\rho_c$ for different values of $\alpha$ is shown in Fig.~\ref{mass2}. It is clear from the figure that $\frac{\partial M}{\partial \rho_c} >0$ everywhere inside the stellar interior for different values of $\alpha$ and therefore our proposed model is stable under Harrison-
Zeldovich-Novikov static stability criterion.

   \section{The generalised Tolman-Oppenheimer-Volkoff equation}\label{5}
For our system, the energy conservation equation of motion is defined by
\[\nabla^{\mu}T_{\mu \nu}=0,\]
   The generalised TOV equation for $f(T)$ gravity can be constructed in the modified form as
   \begin{eqnarray}\label{con1}
-\frac{\nu'}{2}(\rho+p_r)-\frac{dp_r}{dr}+\frac{2}{r}(p_t-p_r)=0,
\end{eqnarray}

   Tolman \cite{Tolman:1939jz} and later Oppenheimer and Volkoff \cite{Oppenheimer:1939ne} proposed that physically realistic models must be stable under the three forces of gravitational force ($F_g$), hydrostatic force ($F_h$), and anisotropic force ($F_a$), such that the sum of the forces becomes zero for the system to be in equilibrium. In other words, $F_g+F_h+F_a=0$. The expression of these three forces are given as follows:

   \begin{eqnarray}
   F_g&=&-\frac{\nu'}{2}(\rho+p_r)=\frac{\alpha B r (a + B + (2 b + a B) r^2 + b B r^4)}{
 4 \pi (1 + a r^2 + b r^4)^2)},\\
 F_h&=&-\frac{dp_r}{dr}=\frac{\alpha r \Big(-a^2 + b + 2 a B - 2 b (a - 2 B) r^2 -
   b^2 r^4\Big)}{4 \pi (1 + a r^2 + b r^4)^2},\\
   F_a&=&\frac{2}{r}(p_t-p_r)=\frac{\alpha r \Big(a^2 - a B + B^2 + a (2 b + B^2) r^2 -
   b (1+2 B r^2 - (b + B^2) r^4)\Big)}{4 \pi (1 + a r^2 + b r^4)^2}
   \end{eqnarray}

   \begin{figure}[htbp]
    \centering
        \includegraphics[scale=.4]{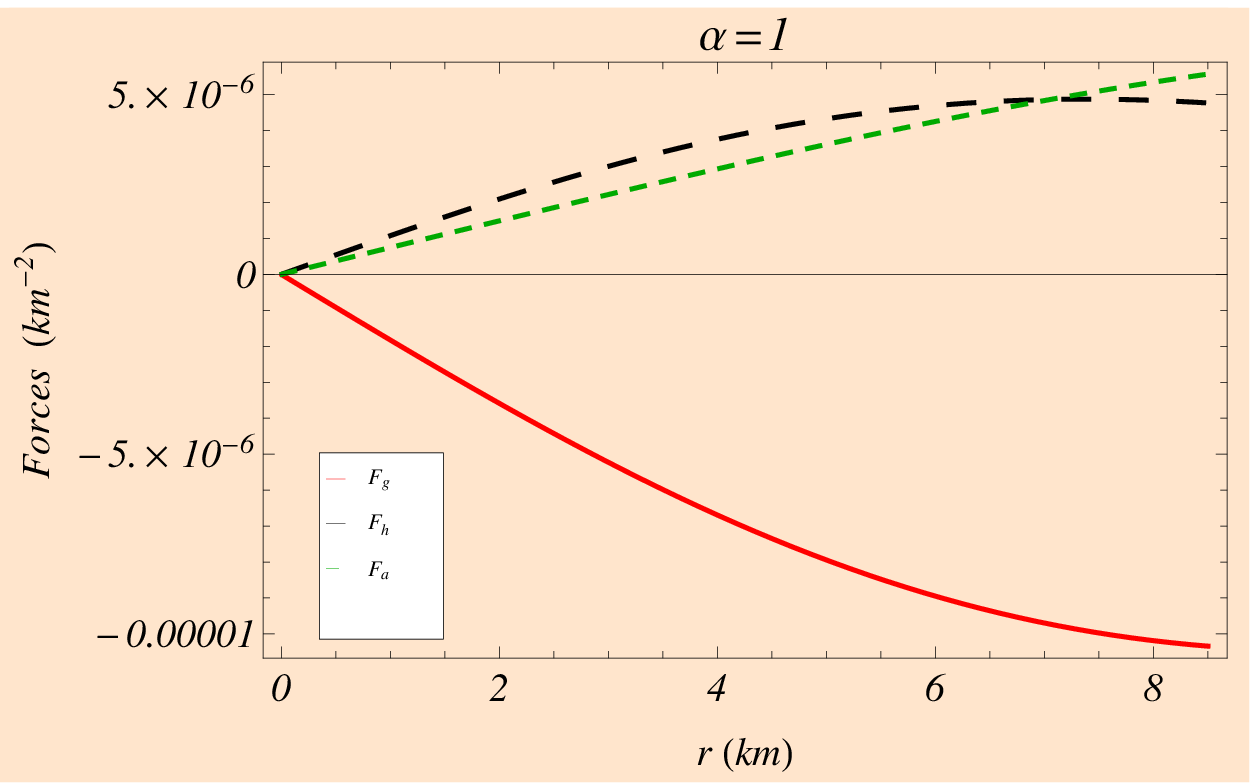}
        \includegraphics[scale=.4]{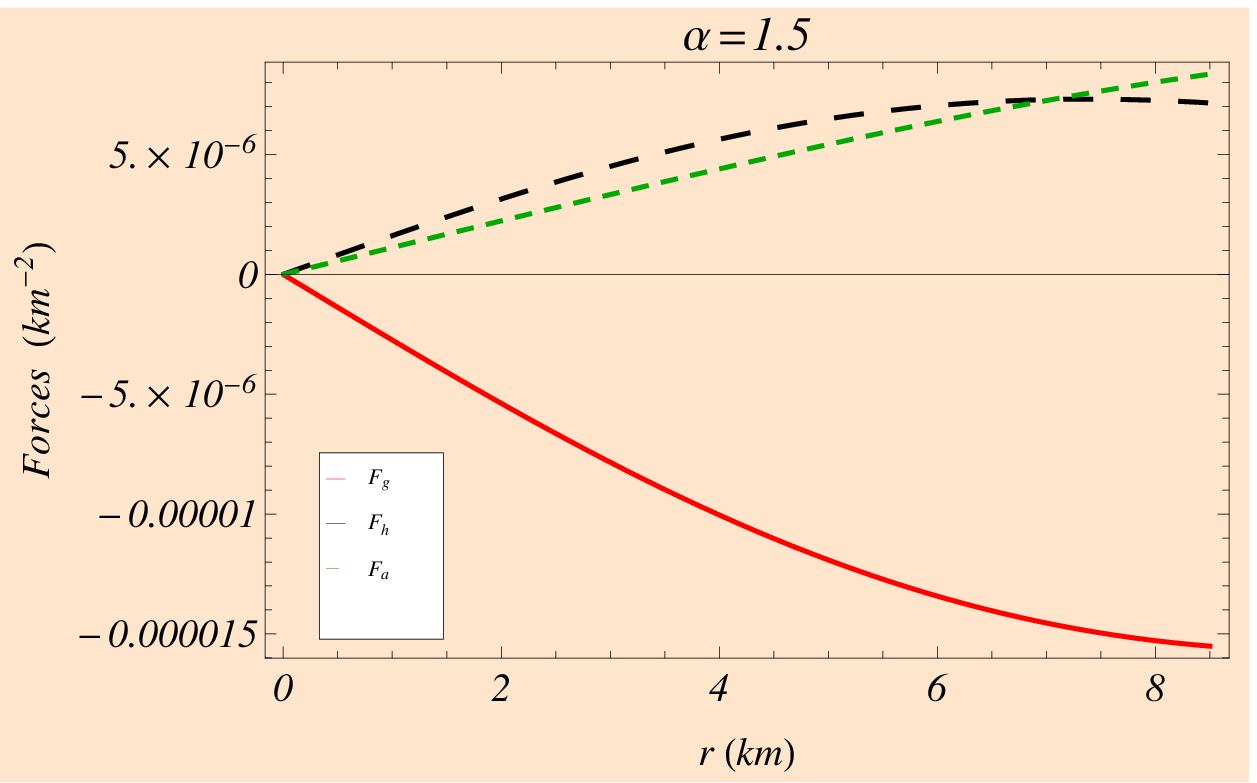}
        \includegraphics[scale=.4]{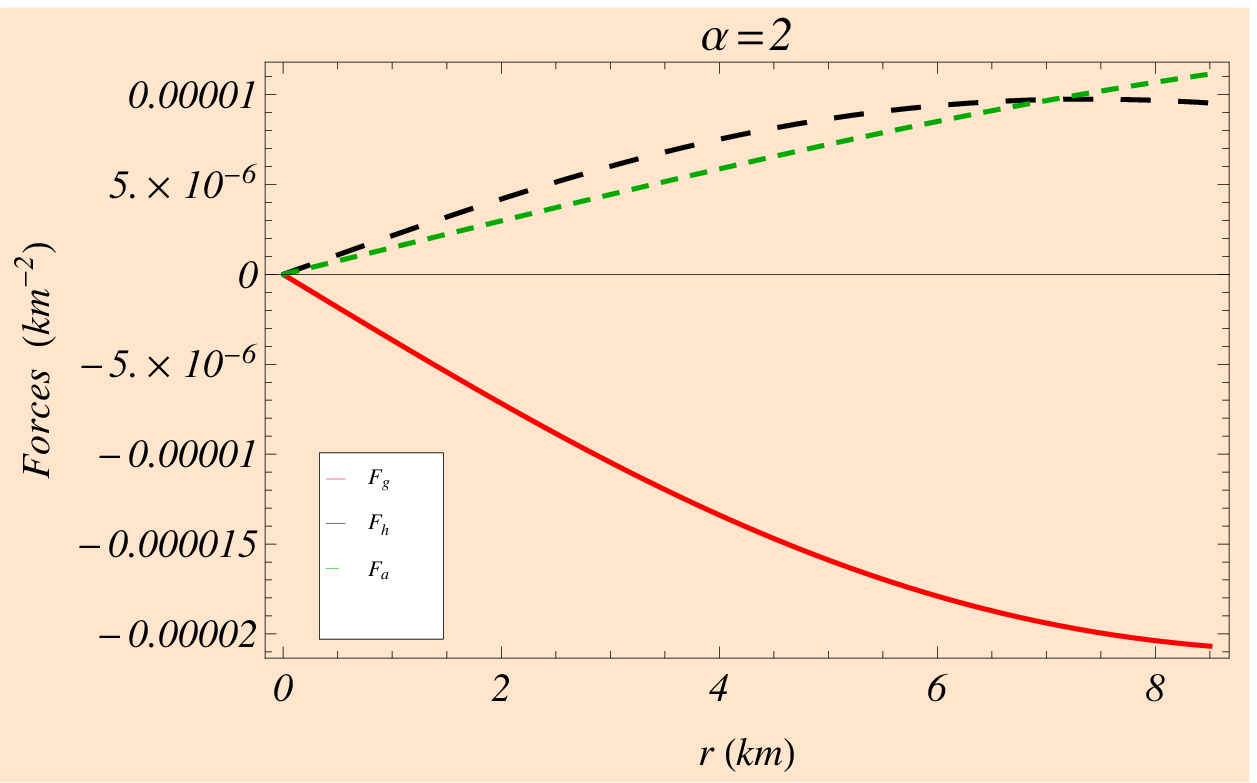}
        \includegraphics[scale=.4]{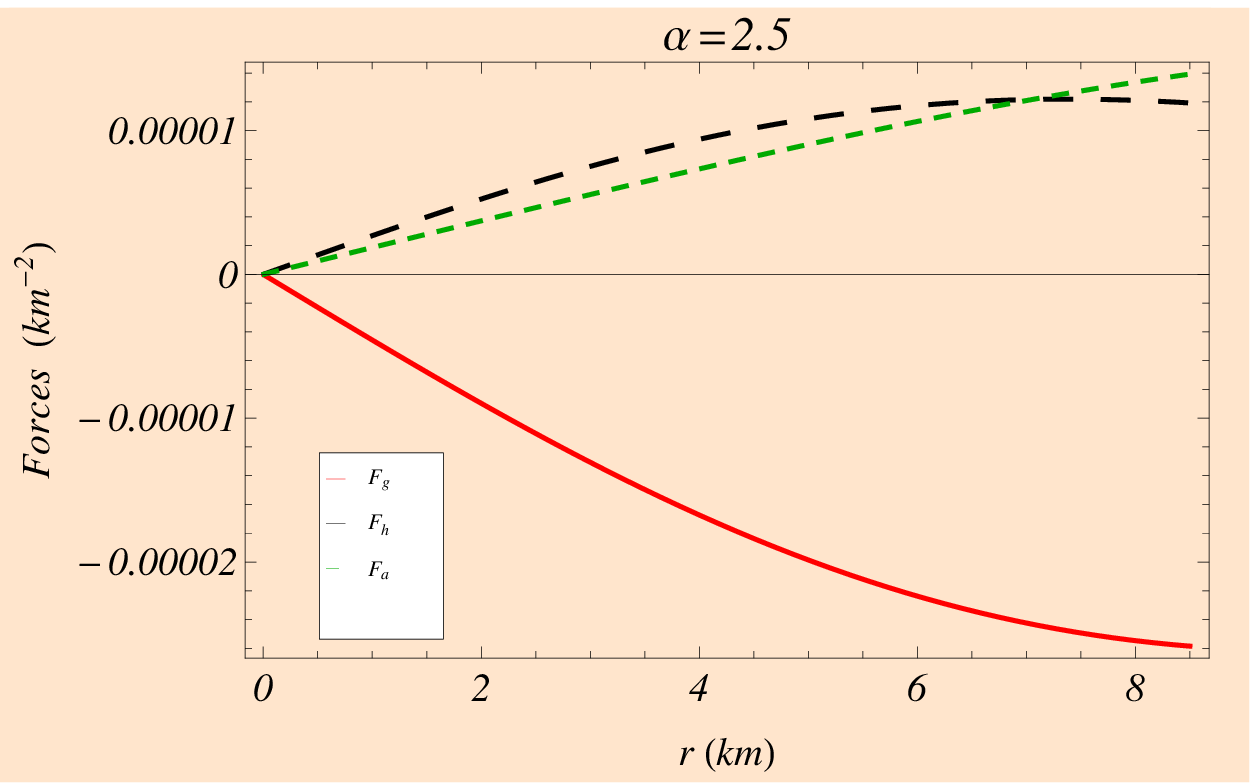}
        \includegraphics[scale=.4]{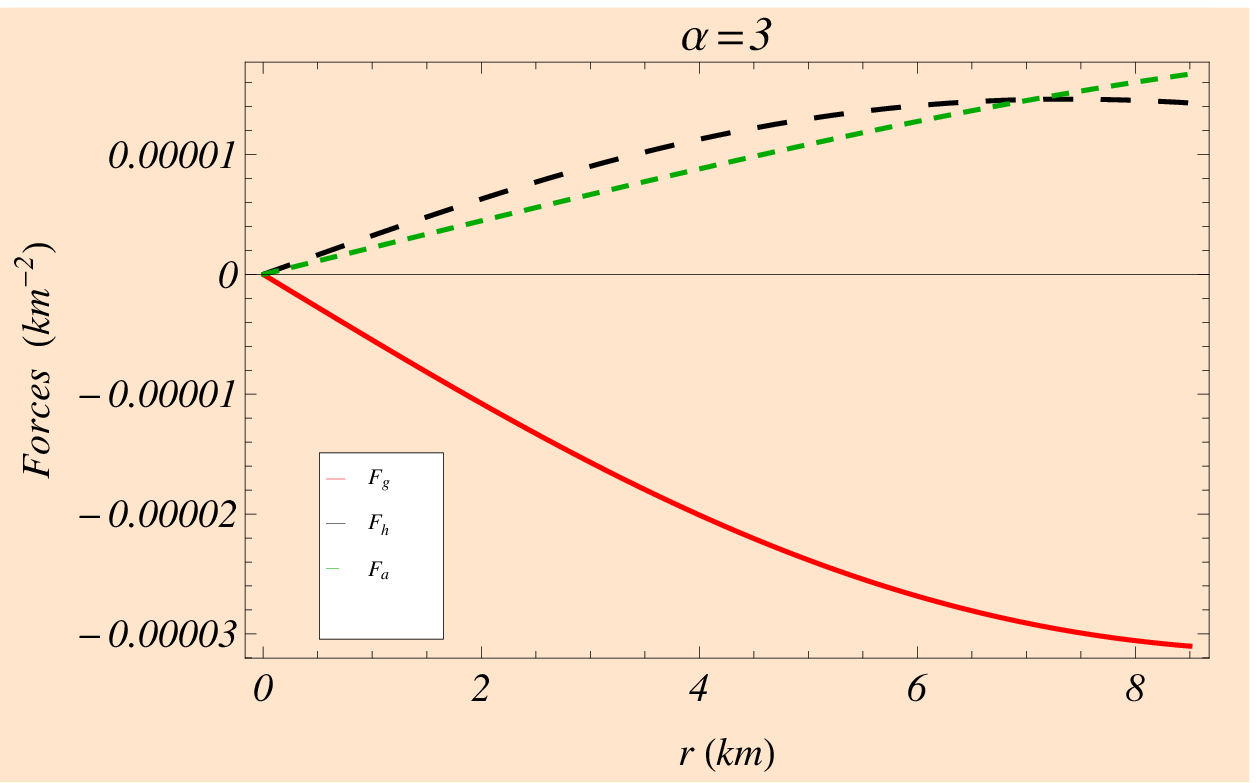}
       \caption{Three different forces acting on the system are shown in the figure for different values of $\alpha$.}
    \label{tov1}
\end{figure}

   The mutual influence of all three forces $F_g$, $F_h$, and $F_a$ clearly justifies the condition of equilibrium for our system, as seen in Fig.~\ref{tov1}.

\section{Conclusion}\label{last}
Over the past few decades, astrophysicists have been actively researching in the creation of super-dense stellar objects. Many characteristics of the compact stellar configuration have been explained by recent improvements in observational techniques but many unsolved issues remain. To explain the internal composition of compact stars and the particle interactions at extremely high densities, theoretical research on compact objects in modified gravity is very necessary. Numerous theoretical concepts have recently been proposed that are consistent with observational data.\par
Theoretical physicists have recently demonstrated an interest in studying the gravitational field as the effect of torsion rather than the curvature of the underlying geometry. The general theory of relativity defines gravity in terms of curvature and it was the basis from which the torsion theory was first developed. As a result, this concept was promoted as teleparallel equivalence of gravity. In the modified TEGR, $f(T)$ gravity has a number of more important characteristics than General Relativity. Recent investigations of orbital motions of the solar system to constrain $f(T)$ gravity have produced some interesting results. This study uses the Tolman-Kuchowicz metric to develop analytical models of compact stars in $f(T)$ gravity with anisotropic matter. Using linear form of $f(T)$ with Tolman-Kuchowicz metric functions, we have determined explicitly the matter components. It has been assumed that the interior region of the star is a static, anisotropic and spherical source. It has shown the possibility to formulate the equations of motion using the diagonal tetrad field. One of the field equations suggests that the unknown function $f(T)$ is a linear function of $T$ with the formula $f(T) = \alpha T + \beta$, where $\alpha$ and $\beta$ are the integration constants. An extensive discussion has been given on the energy, regularity and anisotropy conditions. The unknown constants of the Tolman-Kuchowicz metric have been calculated using the observed values of the mass and radius of the compact star. To draw all the profiles of the model parameters we have chosen the star LMC X-4 with mass $(1.04 \pm 0.09)$  and radius $8.301_{-0.2}^{+0.2}$ km. \cite{Rawls:2011jw}.
The geometry of space and time is also described by the metric potentials. According to the graphical representations, both of the metric potentials satisfy the necessary requirement $e^{\lambda}|_{r=0}=1$ and $e^{\nu}|_{r=0}=D^2$, meaning that the solutions are free from both geometrical and physical singularities. For our present stellar model, the metric potentials $e^{\lambda}$ and $e^{\nu}$ increase both monotonically and nonlinearly from the centre to the surface. The matter density and pressures all are monotonic decreasing functions of `r' and they maintain positive behavior inside the interior of the star. With increasing values of $\alpha$, both the pressures and density take higher values throughout the interior of the star. The first derivatives of density and pressure also indicate that these parameters are maximum at the centre and have a diminishing radial profile. Table~\ref{tbb2} lists the values of central density, surface density, and central pressure. For different values of $\alpha$, the anisotropic force has a repulsive nature. More massive objects are constructed as a result of this repulsive gravitational force. Since the value of $\Delta$ rises as $\alpha$ increases, it is reasonable to infer that the star gets more compact as $\alpha$ increases. In Fig.~\ref{omegar}, the equation of state parameter is shown visually. It can be seen that $\alpha$ has no impact on $\omega_r$ and $\omega_t$ because all plots coincide for different values of $\alpha$. The variation of pressure with respect to the density have been well discussed with the help of graphical representation. The coupling parameter $\alpha$ has no effect on the sound velocity and complexity factor as can be seen in Fig.~\ref{svt}. At the same time the relativistic adiabatic index $\Gamma$ does not depends on $\alpha$ but $\Gamma>4/3$ everywhere inside the stellar interior. The graphical behavior of mass radius relation, compactness and surface redshift function are also shown for different values of $\alpha$ and all are in reasonable range.  Fig.~\ref{ec2} demonstrates that the system we are considering with a suitable choice of mass and radius satisfies all the energy conditions and supports the physical validity of our model. The behavior of the mass function with respect to the central density has been discussed in details. According to profile ~\ref{tov1}, the gravitational force balances out the combined effects of hydrostatic and anisotropic forces, keeping our present super dense compact star in a stable equilibrium position. In this respect we can mention two earlier work done on compact star model in modified gravity. In the $f(G, T)$ theory of gravity, where $G$ and $T$ are the Gauss-Bonnet invariant and trace of stress-energy tensor, respectively, Bhatti et al. \cite{Bhatti:2017fov} investigated several physically realistic aspects for the potential emergence of compact stars. In the presence of an anisotropic source, the authors explain the basic formalism of this modified theory and investigate some relevant aspects utilizing energy conditions and physical parameters. For this systematic examination, three different known star models-namely, SAXJ 1808.43658, Her X-1, and 4U182030-are used. Plots were used to discuss the physical behavior of anisotropic stresses, energy density, energy conditions, measure of anisotropy, and stability of compact stars. In the context of metric $f(R)$ gravity, where $R$ is the Ricci scalar, Yousaf et al.  \cite{Yousaf:2017lto} investigated some plausible configurations of anisotropic spherical structures. The nature of specific compact stars is examined using three different modified gravity models by using metric potential provided by Krori and Barua. In the context of modeling stellar structures, they also describe the behavior of various forces, equation of state parameter, measure of anisotropy and Tolman-Oppenheimer-Volkoff equation and showed that the proposed model in $f(R)$ gravity are feasible and realistic.
As a concluding remark, in this study we have successfully demonstrated that the Tolman-Kuchowicz metric may be used to describe the anisotropic nature of compact stars in a stellar system that is singularity free and completely stable. Here, we found that all physical characteristics of compact stars follow logically consistent patterns, indicating the reliability and validity of the model which have been proposed within the context of modified $f(T)$ gravity. The present work could be extended to find the energy of the particles in the modified gravitational field coming from Einstein-$\Lambda$ gravity.\\

\begin{table*}[t]
\centering
\caption{The numerical values of central density, surface density, central pressure for the compact star LMC X-4 for different values of coupling constant $\alpha$.}
\label{tbb2}
\begin{tabular}{@{}cccccccccccccccc@{}}
\hline
$\alpha$& $\rho_c$ & $\rho_s$ & $p_c$ & $u(R)$ & $z_s(R)$& \\
\hline
 1& $8.16441 \times 10^{14}$& $5.66856\times 10^{14}$ & $3.49106 \times 10^{34}$&0.1475&0.190983\\
 1.5& $1.22466 \times 10^{15}$ & $8.50283 \times 10^{14}$& $5.23659 \times 10^{34}$&0.22125&0.339299\\
 2& $1.63288 \times 10^{15}$ & $1.13371 \times 10^{15}$ & $6.98212 \times 10^{34}$&0.295&0.561738\\
 2.5& $2.0411 \times 10^{15}$ & $1.41714 \times 10^{15}$ & $8.72765 \times 10^{34}$&0.36875&0.9518\\
 3& $2.44932 \times 10^{15}$ & $1.70057 \times 10^{15}$ & $1.04732 \times 10^{35}$&0.4425&1.94884\\
\hline
\end{tabular}
\end{table*}

\section*{Conflict of Interest}There is no conflicts of interest, according to the author, regarding the publication of this article.

\section*{Acknowledgements} P.B. is thankful to the Inter University Centre for Astronomy and Astrophysics (IUCAA), Pune, Government of India, for providing visiting associateship. PB also acknowledges that this work is carried out under the research project Memo No: $649$(Sanc.)/STBT-$11012(26)/23/2019$-ST SEC funded by Department of Higher Education, Science \& Technology and Bio-Technology, Government  of West Bengal.

\bibliography{ft}

\end{document}